\documentclass[pra,twocolumn,aps,floatfix,showpacs,tightenlines,superscriptaddress,amsmath,amssymb]{revtex4}
\usepackage{amssymb,amsbsy,epsfig,color,graphicx}

\begin{document}

\title{Hierarchies of geometric entanglement}

\author{M. Blasone}
\affiliation{Dipartimento di Matematica e Informatica,
Universit\`a degli Studi di Salerno, Via Ponte don Melillo,
I-84084 Fisciano (SA), Italy} \affiliation{INFN Sezione di Napoli,
Gruppo collegato di Salerno, Italy}

\author{F. Dell'Anno}
\affiliation{Dipartimento di Matematica e Informatica,
Universit\`a degli Studi di Salerno, Via Ponte don Melillo,
I-84084 Fisciano (SA), Italy} \affiliation{INFN Sezione di Napoli,
Gruppo collegato di Salerno, Italy}
\affiliation{CNR-INFM Coherentia,
Napoli, Italy; CNISM Unit\`a di Salerno, Italy}

\author{S. De Siena}
\affiliation{Dipartimento di Matematica e Informatica,
Universit\`a degli Studi di Salerno, Via Ponte don Melillo,
I-84084 Fisciano (SA), Italy} \affiliation{INFN Sezione di Napoli,
Gruppo collegato di Salerno, Italy}
\affiliation{CNR-INFM Coherentia,
Napoli, Italy; CNISM Unit\`a di Salerno, Italy}

\author{F. Illuminati}
\thanks{Corresponding author. Electronic
address: illuminati@sa.infn.it}
\affiliation{Dipartimento di Matematica e Informatica,
Universit\`a degli Studi di Salerno, Via Ponte don Melillo,
I-84084 Fisciano (SA), Italy} \affiliation{INFN Sezione di Napoli,
Gruppo collegato di Salerno, Italy}
\affiliation{CNR-INFM Coherentia,
Napoli, Italy; CNISM Unit\`a di Salerno, Italy}
\affiliation{ISI Foundation for Scientific Interchange, Viale
Settimio Severo 65, 00173 Torino, Italy}

\date{May 28, 2008}

\begin{abstract}
We introduce a class of generalized geometric
measures of entanglement. For pure quantum states of $N$
elementary subsystems, they are defined as the
distances from the sets of $K$-separable states ($K=2,\ldots,N$).
The entire set of generalized geometric measures provides
a quantification and hierarchical ordering of
the different bipartite and multipartite components of the global
geometric entanglement, and allows to discriminate among the
different contributions. The extended measures are applied to the
study of entanglement in different classes of $N$-qubit pure states.
These classes include $W$ and $GHZ$ states, and their symmetric
superpositions; symmetric multi-magnon states; cluster states;
and, finally, asymmetric generalized $W$-like superposition states.
We discuss in detail a general method for the explicit
evaluation of the multipartite components of geometric entanglement, and
we show that the entire set of geometric measures establishes an ordering
among the different types of bipartite and multipartite entanglement.
In particular, it determines a consistent
hierarchy between $GHZ$ and $W$ states, clarifying the original result
of Wei and Goldbart that $W$ states possess a larger global entanglement
than $GHZ$ states. Furthermore, we show that all multipartite components
of geometric entanglement in symmetric states obey a property of self-similarity
and scale invariance with the total number of qubits and the number of
qubits per party.
\end{abstract}


\maketitle

\section{Introduction}
\label{Intro} Quantification of pure state bipartite entanglement,
a concept that emerged immediately after the first systematization
of quantum mechanics \cite{Einstein}, is by now well understood in
terms of the entropic content in the reduced states of the
constituent subsystems, as lucidly pointed out for the first time
by Schr\"odinger \cite{Schroedinger}. The universal properties
that any {\it bona fide} measure of entanglement has to satisfy
have been thoroughly discussed and characterized in recent years
\cite{Bennett1,Bennett2,Vidal,PlenioVirmani}.
For pure states of bipartite systems, the von Neumann entropy is
the unique measure of entanglement, and all other consistent
measures are monotonic functions of the former
\cite{Popescu}. However, this uniqueness is lost
in bipartite mixed states: In this context, measures that differ
according to their definitions and/or operational meaning, such
as, for instance, the entanglement of formation, the distillable
entanglement, the relative entropy of entanglement, and the
negativity \cite{Bennett2,EntRelEntr,Negativity}, quantify
different forms of entanglement. In fact, very few of these
quantities can be computed explicitly for mixed quantum states,
even in the simplest instances. A notable
exception is the celebrated Wootters formula for the entanglement
of formation of arbitrary two-qubit mixed states, obtained in
terms of the concurrence \cite{HillWootters,Wootters}.

The situation becomes even more complex in the multipartite
instance, already at the level of pure states in
finite-dimensional Hilbert spaces. Progress has been achieved
mainly in understanding the different ways in which multipartite
systems can be entangled. The intrinsic nonlocal character of
entanglement imposes invariance and monotonicity constraints under
local quantum operations. Equivalence classes of entangled states
can be defined with respect to the group of reversible stochastic
local quantum operations assisted by classical communication
(SLOCC) \cite{SLOCC}. Such an approach has allowed to demonstrate
that three and four qubits can be entangled, respectively, in two
and nine different inequivalent ways
\cite{2diffwayent,9diffwayent}. In the case of three qubits, the
representatives of the two inequivalent classes are, notoriously,
the $W$ and $GHZ$ states \cite{GHZst,2diffwayent}.

Simplifying to the essential, in a multipartite scenario a
legitimate quantification of entanglement can be achieved by
identifying a positive function that is an entanglement monotone
(vanishing on separable states and not increasing under SLOCC),
and is endowed with some kind of operational interpretation.
Several measures satisfying these requirements have been proposed.
For a system of three qubits, Wootters and co-workers defined the
so-called residual entanglement, or $3$-tangle, a quantity
constructed as the difference between the squared three-qubit
concurrence and the squared concurrences of the reduced two-qubit
states \cite{CoffKundWoot}. While successfully detecting the
genuine tripartite entanglement in the state $|GHZ^{(3)}\rangle$,
the $3$-tangle (or residual tangle) vanishes if computed for the
state $|W^{(3)}\rangle$, thus being not appropriate for the
quantification of tripartite entanglement in this class of states.
In other words, a non vanishing residual tangle is a sufficient
but not necessary condition for the detection of genuine
multipartite entanglement. The Schmidt measure, defined as the
minimum of $\log_{2} r$ with $r$ being the minimum of the number
of terms in an expansion of a quantum state in product basis, has
been proposed by Eisert and Briegel as an alternative measure of
multipartite entanglement \cite{EisertBriegel}. Other proposals
are given as functions of the various bipartite entanglements
contained in a multipartite state
\cite{Wallach,Brennen,Scott,Oliveira,Pascazio}. The seed
representative of this class of measures is the global
entanglement of Meyer and Wallach, that for an $N$-qubit state is
defined as the sum of all the possible two-qubit concurrences
\cite{Wallach}.

A different set of entanglement quantifiers is defined in purely
geometric terms. The relative entropy of entanglement (generalized
for multipartite settings) and the so-called geometric
entanglement belong to this class
\cite{multirelentropy,GMEShimony,GMEBarnumLind}. The relative
entropy of entanglement is defined as the distance of a
given state from the set of fully separated states, quantified in
terms of the quantum relative entropy \cite{multirelentropy}. The
geometric entanglement was originally defined as the Euclidean
distance of a given multipartite state to the nearest fully
separable state \cite{GMEShimony,GMEBarnumLind,GMEWeiGold}. This
last measure can be considered as one of the most reliable
quantifiers of global multi-particle entanglement \cite{EisertGross}: It
exhibits interesting connections with other measures
\cite{GMEWeiGold,Cavalcanti} and can be efficiently estimated
by quantitative entanglement witnesses amenable of experimental
verification \cite{Guehne,Eisert}. Given an $N$-partite pure state $|\Psi\rangle$, 
the geometric measure of entanglement introduced by Wei and Goldbart \cite{GMEWeiGold}
is defined as:
\begin{equation}
E_{G}(|\Psi\rangle) = 1 - \max_{|\Phi\rangle} \,
\Big|\langle \Phi|\Psi\rangle\Big|^{2} \; ,
\label{OriginalMeasure}
\end{equation}
where the maximum is taken with respect to all pure states
that are fully factorized, i.e. the $N$-separable states
\begin{equation}
|\Phi\rangle \, = \, \bigotimes_{s=1}^{N} \,
|\Phi_{s}\rangle \; ,
\label{NseparablePhi}
\end{equation}
where the states $|\Phi_{s}\rangle$ are single-qubit pure states.
This measure is intrinsically geometric because it coincides
with the distance (in the Hilbert-Schmidt norm) between a
given pure state and the set of fully separable (i.e. fully product) pure
states. The Wei-Goldbart geometric measure is thus a global quantifier of
entanglement, including all the bipartite and multipartite contributions.

The geometric measure can be extended by the convex roof procedure to the
case of mixed states, and, analogously to the Meyer-Wallach global entanglement,
is a proper multipartite entanglement monotone.
Remarkably, the geometric measure can be effectively exploited
to quantify the entanglement of two distinct multipartite bound
entangled states \cite{GMEBoundEntang}
and to study the behavior of global entanglement at the approach of
quantum phase transitions \cite{GMEQpts,GMEOrus,GMELmg}.
%%Recently, generalized versions of this measure have been proposed \cite{GMEGeneralizations}.
%%Furthermore, procedures for the analytical determination of the geometric measure
%%have been presented \cite{GMEMethods}.
However, notwithstanding the very appealing properties and
the important results cited above, the global nature of the Wei-Goldbart
geometric entanglement constitutes a limitation insofar as it does not allow
to distinguish and discriminate among the different bipartite and multipartite
contributions to the overall entanglement, to determine their properties,
and to establish a systematic hierarchy among them. It is the aim of the
present work to fill this gap. \\
In this paper, we define and study in detail a natural and powerful
multipartite generalization of the
geometric measure of entanglement for pure states of many-qubit systems.
We first introduce a compact and convenient parametrization to express
analytically general $K$-separable states of $N$-qubit systems ($K
\leq N$). We then analyze the behavior of the distance between pure $N$-qubit
states and the set of $K$-separable states ($K=2,\ldots,N$) in order
to determine and distinguish the different multipartite contributions to
the geometric entanglement and characterize their ordering. The different
distances, corresponding to $K=2,\ldots,N$, quantify hierarchically the different
forms of multipartite entanglement present in the given $N$-qubit state. In Section
\ref{secGME}, we define the multi-component generalization of the geometric
measure, we review the known results in the case of full
separability and, for this latter case, we also present some
further extended results. In Section \ref{secKsepGME} we evaluate
explicitly the generalized multi-component geometric measure of entanglement,
considering genuine $K$-separability ($K \leq N$). We analyze the detailed behavior
of the different forms of geometric entanglement for various relevant classes of
$N$-qubit states, establishing some generic and asymptotic properties, and we determine
the explicit hierarchy holding for $W$, $GHZ$, cluster, and multi-magnon states.
In the case of $W$ and $GHZ$ states, the established relations
between the different forms of multipartite geometric
entanglement clarify the original result of Wei and Goldbart
that $W$ states possess a larger total entanglement
content than $GHZ$ states, when quantified by the geometric measure.
Moreover, in the case of $N$-qubit $W$ states, we find that the
geometric entanglement is scale-invariant (self-similar) as
the total number of qubits grows at the same rate as the number
of subsystems in each party.
We show that the property of self-similarity is enjoyed by other symmetric states
as well, as a direct consequence of the invariance under permutation of any two qubits,
which is the characterizing property of symmetric states.
We then analyze and determine the different multipartite components of
geometric entanglement for arbitrary symmetric superpositions of $N$-qubit $GHZ$ and $W$ states.
In Section \ref{SecGeneralW} we compute the multipartite geometric measures
for classes of generalized $W$ states beyond the single-excitation regime:
these are general symmetric states, the so-called magnon states, that are
of crucial importance, for instance, in the theory of magnetism.
Furthermore, we determine the multipartite components of geometric entanglement
in generalized, asymmetric $W$-like superposition states.
In this and related cases, at variance with most of the symmetric instances, a complete
characterization of entanglement requires the determination of all
the multipartite geometric components, which we compute explicitly.
Finally, in Section \ref{Conclusions} we discuss some general
conjectures on generic properties and typical behaviors of the geometric entanglement,
and examine some outlooks on possible future lines of research.

\section{geometric entanglement: $K$-separability vs. full separability}
\label{secGME}

Let us consider a $N$-qubit system, corresponding to a tensor-product
state space $\mathcal{H}^{d_{N}}$ of dimension $d_{N} = 2^{N}$.
For such a system, let us introduce the integer $K$, $2 \leq K \leq N$, and the
ordered sequence of integers $\{ M_1,M_2,\ldots,M_K \}$, where
$M_1 \leq M_2 \leq \ldots \leq M_K$, and $\sum_{s=1}^{K} M_s = N$.
Let us consider the $K$-partition of the system in $K$
subsystems described by the sets $\{Q_{s}\}_{s=1}^{K}$.
Let each set $Q_{s}$ be composed of $M_{s}$ elementary parties, i.e.
$Q_{s}=\{ i_{1}^{(s)},i_{2}^{(s)},\ldots, i_{M_{s}}^{(s)} \}$,
where $i_{j}^{(s)} \in \{1,\ldots,N\}$ is a discrete index labeling the $N$
elementary parties, and $Q_{s}\bigcap Q_{s'} = \emptyset$ for $s \neq s'$.
Given a generic $K$-partition $Q_1|Q_2|\ldots|Q_K$ of the $N$-qubit system,
any $K$-separable state associated to such a partition is defined as
the tensor product of $K$ $M_{s}$-qubit
pure states $|\Phi_{s}^{(Q_{s})}\rangle$. Each state $|\Phi^{(Q_{s})}\rangle$
belongs to the Hilbert space $\mathcal{H}^{d_{Q_{s}}}$ of dimension
$d_{Q_{s}} = 2^{M_{s}}$.
A $K$-separable state can then be written as
\begin{equation}
\bigotimes_{s=1}^{K} \,
|\Phi_{s}^{(Q_{s})}\rangle \; .
\label{KseparablePhi}
\end{equation}
Correspondingly, the Hilbert space $\mathcal{H}^{d_{N}}$ is decomposed in the tensor product
$\bigotimes_{s=1}^{K}\mathcal{H}^{d_{Q_{s}}}$.
Varying the integers $M_s$,
one obtains different $K$-partitions $Q_1|Q_2|\ldots|Q_K$ and,
correspondingly, different possible $K$-separable states.
It is worth noticing that even at fixed $M_1|M_2|\ldots|M_K$,
there exist different $K$-partitions associated with the different arrangements
of the elementary parties in the sets $Q_{s}$;
in fact, from a given initial $K$-partition, a certain number of different $K$-partitions
can be generated through permutations of the elementary parties
belonging to different sets $Q_{s}$.
We then denote by ${\mathcal{\mathbf{S}}}_K$ the set of all $K$-separable states,
defined as
\begin{equation}
{\mathcal{\mathbf{S}}}_K = \bigcup_{\{Q_1,\ldots,Q_K\}} S_K (Q_1|\ldots|Q_K) \; ,
\label{CheClass}
\end{equation}
where $S_K (Q_1|Q_2|\ldots|Q_K)$ is the set of all the $K$-separable
states associated to a fixed $K$-partition.
We can now define the relative (i.e. partition-dependent)
and the absolute (i.e. partition-independent) geometric
measures of entanglement with respect to $K$-separable pure states for
an arbitrary $N$-qubit pure state $|\Psi^{(N)}\rangle$, respectively, as:
\begin{equation}
E_{G}^{(K)}(Q_1 |\ldots |Q_K ) \,=\, 1-\Lambda_{K}^{2}(Q_1 |\ldots |Q_K ) \,,
\label{GMErel}
\end{equation}
where the squared overlap
\begin{equation}
\Lambda_{K}^{2}(Q_1 |\ldots |Q_K ) \,=\, \max_{|\varphi\rangle \in
S_{K}(Q_1 |\ldots |Q_K ) } \,
\Big|\langle \varphi| \Psi^{(N)}\rangle\Big|^{2} \; ,
\label{overlap2rel}
\end{equation}
and
\begin{equation}
E_{G}^{(K)}(|\Psi^{(N)}\rangle) \,=\, 1-\Lambda_{K}^{2}(|\Psi^{(N)}\rangle) \,,
\label{GMEabs}
\end{equation}
where the squared overlap
\begin{equation}
\Lambda_{K}^{2}(|\Psi^{(N)}\rangle) \,=\, \max_{|\Phi\rangle \in
{\mathcal{\mathbf{S}}}_{K}} \,
\Big|\langle \Phi|\Psi^{(N)}\rangle\Big|^{2} \; .
\label{overlap2abs}
\end{equation}
By Eqs.~(\ref{GMErel}), (\ref{overlap2rel}), (\ref{overlap2abs}),
the quantity (\ref{GMEabs}) measures the
absolute minimum distance of a state from the set of all $K$-separable
states. Equivalently,
$E_{G}^{(K)}(|\Psi^{(N)}\rangle)=\min_{\{S_{K}(Q_1 |\ldots |Q_K )\}}E_{G}^{(K)}(Q_1 |\ldots |Q_K )$.
Trivially, for any $N$-partition (i.e. $K=N$), one has $M_1=M_2=\ldots=M_N=1$
and $N$-separability coincides with full separability, while $1$-separability is
a common feature of any state, i.e. $E_{G}^{(1)} = 0$ for all states $\{
|\Psi^{(N)}\rangle \}$.
In the particular instance of symmetric states,
that are states invariant under the permutation of any two qubits,
the quantities $E_{G}^{(K)}(Q_1 |\ldots |Q_K )$ and $\Lambda_{K}^{2}(Q_1 |\ldots |Q_K )$
satisfy the same invariance property.
Therefore, in such a case, in the definitions (\ref{GMErel}) and (\ref{overlap2rel}),
the symbols $Q_{s}$ can be replaced by the indices $M_{s}$
as the multi-index $M_1 |\ldots |M_K$ completely determines
the particular component of geometric entanglement.
In Ref. \cite{GMEWeiGold}, the measure
(\ref{GMEabs}) is defined only in the simplest instance of $N$-separability. In this
case, we may write the general expression for a (normalized) $K$-separable
state $|\Phi\rangle$, Eq.~(\ref{NseparablePhi}), in the following Hartree form:
\begin{equation}
|\Phi\rangle \,=\, \bigotimes_{l=1}^{N} \Big(\cos\Gamma_{l}
|0\rangle_{l} + e^{i\Delta_{l}} \sin\Gamma_{l} |1\rangle_{l}\Big) \; ,
\label{Nsepstate}
\end{equation}
with $\Gamma_{l}$ and $\Delta_{l}$ real. By using
Eq.~(\ref{Nsepstate}) with $\Delta_{l}=0$, the geometric
measure of entanglement can be analytically computed for
the classes of $GHZ$ and $W$ states. The definition of these
states for the $N$-qubit case reads:
\begin{eqnarray}
&&|GHZ^{(N)}\rangle \,=\, \frac{1}{\sqrt{2}} \sum_{i=1}^{2} \,
|\delta_{i,2}\delta_{i,2}\ldots\delta_{i,2}\rangle \,,
\label{GHZN} \\
&& \nonumber \\
&&|W^{(N)}\rangle \,=\, \frac{1}{\sqrt{N}} \sum_{i=1}^{N} \,
|\delta_{i,1}\delta_{i,2}\ldots\delta_{i,N}\rangle \,,
\label{WN}
\end{eqnarray}
where $\delta_{i,j}$ denotes the Kronecker delta, and $|e^{(1)}
e^{(2)} \ldots e^{(N)}\rangle \,\equiv \, |e^{(1)}\rangle_{1}
|e^{(2)}\rangle_{2} \cdots |e^{(N)}\rangle_{N}$ $(e^{(j)}=0,1)$.
The $GHZ$ and $W$ states are fully symmetric, i.e. invariant under
the exchange of any two qubits, and greatly differ from each other
in their correlations properties. On general grounds \cite{EisertBriegel},
one can expect that $N$-qubit $GHZ$ states must possess $N$-partite entanglement
but no $K$-partite one for $K < N$. On the other hand, the
$N$-qubit $W$ states do possess $K$-partite entanglement for $K < N$.

For the total geometric entanglement of states $|GHZ^{(N)}\rangle$
and $|W^{(N)}\rangle$, measured with respect to the set of
$N$-separable (i.e. fully separable) states, the following
relations hold \cite{GMEWeiGold}:
\begin{eqnarray}
&&\Lambda_{N}^{2}(|GHZ^{(N)}\rangle) \,=\, \frac{1}{2} \,, \label{GMENGHZ}
\\ && \nonumber \\
&&\Lambda_{N}^{2}(|W^{(N)}\rangle) \,=\, \left(\frac{N-1}{N}\right)^{N-1}
\,. \label{GMENW}
\end{eqnarray}
In particular, Eq.~(\ref{GMENW}) is obtained by setting
$\Gamma_{l} = \arcsin (1/\sqrt{N})$, with $l=1,\ldots,N$.
Therefore, for the $|GHZ^{(N)}\rangle$ states, the total geometric entanglement takes
the constant value $1/2$, independently from $N$. On the other hand,
for the $|W^{(N)}\rangle$ states, the total geometric entanglement grows with $N$,
converging to a simple function of the Neper number in the asymptotic limit:
\begin{eqnarray}
&& E_{G}^{(3)}(|W^{(3)}\rangle) \,=\, \frac{5}{9} \approx 0.555 \; , \\
&& E_{G}^{(4)}(|W^{(4)}\rangle) \,=\, \frac{37}{64} \approx 0.578 \; , \\
&& \ldots \nonumber  \\
&& E_{G}^{(N)}(|W^{(N)}\rangle) \,=\, 1 - \left( \frac{N-1}{N} \right)^{N-1} \; , \\
&& \ldots \nonumber  \\
&& \lim_{N\rightarrow\infty}E_{G}^{(N)}(|W^{(N)}\rangle) \,=\, 1-e^{-1}
\approx 0.632 \; .
\end{eqnarray}
Therefore, according to the measure of total geometric entanglement,
the $W$ states are overall more entangled than $GHZ$ states for any $N$,
notwithstanding the fact that the latter must always possess a larger amount
of genuine $N$-partite entanglement. Moreover, the asymptotic limit acquired
by the total geometric entanglement on $W$ states for large $N$ appears to point
at some underlying topological structure.

\begin{figure}[h]
\centering
\includegraphics*[width=8cm]{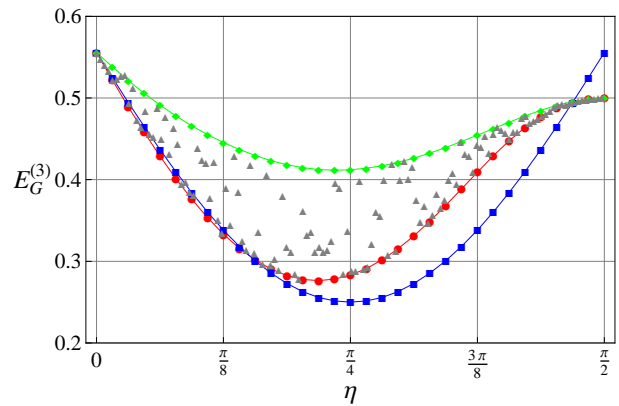}
\caption{(Color online) $E_{G}^{(3)}$ for the superposition of $|W^{(3)}\rangle$
and $|\tilde{W}^{(3)}\rangle$ states, Eq.~(\ref{superposWWt}), and for
the superposition of $|W^{(3)}\rangle$ and $|GHZ^{(3)}\rangle$ states,
Eq.~(\ref{superposWGHZ}), as a function of the mixing angle $\eta$. $E_{G}^{(3)}$
for the state (\ref{superposWGHZ}) is plotted for the following
choices of the free relative phase $\phi$: $\phi=0$ (round points, in red), $\phi=\pi$
(diamond points, in green), and $\phi$ taking random values in the range
$[0,\pi]$ (triangle points, in grey). $E_{G}^{(3)}$ for the state
(\ref{superposWWt}) does not depend on $\phi$ (box points, in blue).
All plotted quantities are dimensionless.}
\label{GMESuperposWG}
\end{figure}

In the first nontrivial multipartite case $N=3$, interesting results have been obtained
also for superposition states of the form \cite{GMEWeiGold}:
\begin{eqnarray}
&&|W\,\tilde{W}^{(3)}\rangle \,=\, \cos\eta |W^{(3)}\rangle +
e^{i\phi}\sin\eta  |\tilde{W}^{(3)}\rangle \,,
\label{superposWWt}
\\ && \nonumber \\
&&|W\,GHZ^{(3)}\rangle \,=\, \cos\eta |W^{(3)}\rangle +
 e^{i\phi}\sin\eta |GHZ^{(3)}\rangle \,, \nonumber \\ &&
 \label{superposWGHZ}
\end{eqnarray}
where the mixing angle $\eta$ lies in the range
$\left[0,\frac{\pi}{2}\right]$, $\phi$ is a free relative phase, and
$|\tilde{W}^{(3)}\rangle \,=\, \frac{1}{\sqrt{3}}(|110\rangle +
|101\rangle + |011\rangle)$. The geometric entanglement is computed with respect to
the fully three-separable state (Eq. \ref{Nsepstate} with $N=3$).
 In Fig. \ref{GMESuperposWG}, $E_{G}^{(3)}$ for the states
(\ref{superposWWt}) and (\ref{superposWGHZ}) is plotted as a
function of $\eta$.
The geometric measure of entanglement for the state (\ref{superposWWt}) attains
its maximum $5/9$ at $\eta=0,\pi/2$ and its minimum at $\eta=\pi/4$, and is
independent of the phase $\phi$; on the contrary, for the
state (\ref{superposWGHZ}) it exhibits an explicit dependence on
$\phi$ that is maximized for $\phi=\pi$ and attains its
maximum value $5/9$ at $\eta=0$. The free relative phase $\phi$ cannot
be eliminated by local unitary operations (in the sense of being of dimension
less than $N$) for the states of the form (\ref{superposWGHZ}), but only by
means of global $N$-dimensional transformations. Therefore, the global
entanglement content of these states must necessarily depend on $\phi$, and the latter
thus acquires the meaning of a global geometric phase.

\section{Multipartite components of geometric entanglement}

\label{secKsepGME}

As discussed above, the distance of a $N$-partite state $|\Psi^{(N)}\rangle$
from the set of fully separable (i.e. $N$-separable) states is a legitimate
quantifier of a global form of entanglement, encompassing $N$-partite,
($N-1$)-partite, $\ldots$, and bi-partite components in an indistinguishable
way. This observation motivates the search for a more refined geometric
quantification of entanglement, in order to distinguish the different multipartite
contributions. To this end, we proceed to study the distances of $|\Psi^{(N)}\rangle$
from the various sets of $K$-separable states, as defined in the previous section.
For a fixed $K$ ($K=2,\ldots,N$), the distance Eq. (\ref{GMEabs}) quantifies the
$N$-, $\ldots$, $(N - K + 2)$-partite contributions to the global entanglement.
Moreover, it is evident that, for each $K$,
\begin{equation}
{\mathcal{\mathbf{S}}}_{K-1} \supseteq {\mathcal{\mathbf{S}}}_{K} \, , \quad
E_{G}^{(K - 1)}(|\Psi^{(N)}\rangle) \leq E_{G}^{(K)}(|\Psi^{(N)}\rangle) \, ,
\label{MatchPoint}
\end{equation}
where the second inequality follows by the law of set inclusion.
Some simple examples may be of help to elucidate the structure of this hierarchy. Let us
take $N=3$. In this case, we have two possibilities: $K=2,3$. For $K=2$ one has
information only on the pure three-partite (three-qubit) component of the geometric entanglement,
while for $K=3$ (distance from the fully separable states) one has undistinguishable
information on both three- and two-qubit entanglement. Moreover, as already mentioned above,
since the set of biseparable states $S_2(1|2)$ contains the set of three-separable
states $S_3(1|1|1)$, it follows that $E_{G}^{(2)}(|\Psi^{(3)}\rangle)
\leq E_{G}^{(3)}(|\Psi^{(3)}\rangle)$. If equality holds,
it then follows that the entire content of entanglement is due only to the three-partite contribution.
The extension to higher dimensions $N \geq 4$ is straightforward, although the number of possible
partitions quickly grows. On the other hand, we will show that the genuine $N$-partite entanglement
of $GHZ$ and $W$ states is always associated to the distance from the set of biseparable states
$S_2(1|N-1)$.

%%Clearly, for a fixed $K \neq N$, there will be inequivalent
%%$K$-separable qubit states $|\Phi\rangle$ (belonging to the same set)
%%that correspond to the different inequivalent ways of $K$-partitioning
%%the $N$-partite system. Let us indeed consider the $K$-partition of the
%%$N$-partite system $P=\{P_{1},P_{2},\ldots,P_{N}\}$ in $K$
%%subsystems $\{Q_{s}\}$, where $s = 1, \ldots, K$. Each subsystem $Q_{s}$ is
%%composed of $n_{s}$ parties $\{P_{i_{s}}\}$, where $i_{s}
%%\in \{1,\ldots,N\}$, $\sum_{s}n_{s} = N$, and  $Q_{s}\bigcap
%%Q_{s'} = \emptyset$ $(i_{s} \neq i_{s'})$. For a given
%%$K$-partition, there exist $F(n_{1},\ldots,n_{K})$ different, inequivalent
%%ways to $K$-partite the system, with
%%\begin{equation}
%%F(n_{1}, \ldots, n_{K})=\frac{N!}{n_{1}!\, n_{2}! \cdots n_{K}!
%%\,d!} \,, \quad\; d=\sum_{s\neq s'=1}^{K}\delta_{n_{s},n_{s'}}.
%%\label{Kdecompnumber}
%%\end{equation}
%%The factor $d$ is an integer related to the degeneracy in the
%%integers $n_{s}$. In this work we will consider only symmetric states,
%%invariant under the permutation of any two parties.  of states
%%allows a great simplification as, at fixed $K$-partition, the GME
%%is invariant
%%under permutation of any parties. \\

We shall now introduce some concise notations that will be useful
in the following. Let us denote by $|\chi^{(M)}\rangle$ an arbitrary
$M$-partite qubit state, that can be
expressed in the form
\begin{equation}
|\chi^{(M)}\rangle = \sum_{j_{1},\ldots,j_{M} = 0}^{1}
c_{j_{1},\ldots,j_{M}}|j_{1} \ldots j_{M} \rangle \; ,
\label{chi1}
\end{equation}
where $c_{j_{1},\ldots,j_{M}}$ are complex parameters satisfying
the normalization constraint $\sum_{j_{1},\ldots,j_{M} = 0}^{1}
|c_{j_{1},\ldots,j_{M}}|^{2} = 1$. In order to simplify the
notation, we substitute the multi-index
$(j_{1},\ldots,j_{M})$ by the single index $J = \sum_{q=1}^{M}
 2^{M-q} \, j_{q}$ (i.e. summation in the binary system),
so that Eq.~(\ref{chi1}) reads
\begin{equation}
|\chi^{(M)}\rangle = \sum_{J =
0}^{d_{M}-1} c_{J} | J \rangle\rangle_{M} \; .
\label{chi}
\end{equation}
Obviously, one has $0 \leq J \leq d_{M}-1=2^{M}-1$.
This notation
provides a useful ordering of the states based on the binary
numbering. In fact, the index $J=0,1,2 \ldots, 2^{M}-1$ labels,
respectively, the states
$|00 \ldots 00 \rangle, |00 \ldots 01 \rangle, |00 \ldots 10 \rangle, \ldots,
|11 \ldots 11 \rangle$.
Let us note that the states $|J\rangle\rangle_{M}$ satisfy the orthonormality relation,
i.e. $\,_{M}\langle\langle J|J' \rangle\rangle_{M} = \delta_{J,J'}$;
moreover, each $M$-qubit state $|J\rangle\rangle_{M}$ can be written in the
decomposed form
$|J\rangle\rangle_{M} = |J_{1}\rangle\rangle_{M_{1}} \otimes |J_{2}\rangle\rangle_{M_{2}}$,
where $M=M_{1}+M_{2}$, and $J= 2^{M_{2}} J_{1}  + J_{2}$.
Using the Euler representation and eliminating an irrelevant
global phase factor, the parameters $c_{J}$ can be cast in the form $c_{J} =
r_{J} e^{i \phi_{J}}$, where $r_{J}=|c_{J}|$, $\phi_{0}=0$, and the phases
$\phi_{J}$ are arbitrary for $J > 1$.
It is worth noting that the fully separable
state (\ref{Nsepstate}) is a particular realization of
$|\chi^{(M)}\rangle$ for $M=N$.
The $K$-separable state given by Eq.~(\ref{KseparablePhi}), can be
expressed explicitly by using, for each state
$|\Phi_{s}^{(M_{s})}\rangle$, the general form (\ref{chi}) and
the hyperspherical parametrization
introduced in Appendix \ref{AppHypPar}, see Eq.~(\ref{hypersphparam}).
The hyperspherical parametrization will then prove extremely convenient
in the computation of Eq.~(\ref{overlap2abs}) for any value of the index $K$.
By using the notation in terms of the binary-numbering index,
Eqs.~(\ref{GHZN}) and (\ref{WN}) can be recast as:
\begin{eqnarray}
&& |GHZ^{(N)}\rangle \,=\, \frac{1}{\sqrt{2}}(|0\rangle\rangle_{N} + |2^{N}-1\rangle\rangle_{N}) \,,
\label{GHZNbin} \\
&& \nonumber \\
&& |W^{(N)}\rangle \,=\, \frac{1}{\sqrt{N}} \sum_{p=0}^{N-1} \, |2^{p}\rangle\rangle_{N} \,.
\label{WNbin}
\end{eqnarray}

In the next subsections we will determine the different multipartite
contributions for some relevant classes of states symmetric under
exchange of any pair of qubits.

\subsection{Three-qubit pure states}
\label{subsecA}

We begin by considering three-qubit pure states, the simplest nontrivial
instance of multipartite states. In this case, given the tensor product
Hilbert space $\mathcal{H}^{(8)} = \mathcal{H}^{(2)} \otimes \mathcal{H}^{(2)}
\otimes \mathcal{H}^{(2)}$, associated to a system of $N = 3$ qubits, there
are only two sets of separable states: The set ${\mathcal{\mathbf{S}}}_2$
of biseparable states ($K = 2$), and the set ${\mathcal{\mathbf{S}}}_3$ of
three-separable states ($K = 3$, full separability), with
${\mathcal{\mathbf{S}}}_2 \supseteq {\mathcal{\mathbf{S}}}_3$. The distance $E_{G}^{(3)}$ from
the set ${\mathcal{\mathbf{S}}}_3$ measures the global geometric entanglement of Wei and Goldbart,
while the distance $E_{G}^{(2)}$ from the set $S_2$ measures the genuine three-partite
contribution to the global geometric entanglement: $E_{G}^{(2)} \leq E_{G}^{(3)}$, with
equality holding when all the entanglement is due only to the genuine tripartite
component and there is no bipartite component. The general expression for
any biseparable state $|\Phi\rangle$
is of the form:
\begin{equation}
|\Phi\rangle = |\Phi_{1}^{(1)}\rangle_{k} \otimes |\Phi_{2}^{(2)}\rangle_{ij} \; ,
\label{2sep3qubitstate}
\end{equation}
where
\begin{eqnarray}
& & |\Phi_{1}^{(1)}\rangle = \Big(\cos\Gamma |0\rangle +
e^{i\Delta} \sin\Gamma |1 \rangle \Big) \; , \nonumber \\
&& \label{Phi12} \\
& & |\Phi_{2}^{(2)} \rangle = \Big(\cos\delta_{1} |00 \rangle +
e^{i\phi_{2}}\sin\delta_{1}\cos\delta_{2} |01 \rangle +
\nonumber \\
&& {}\hspace{-.5cm} e^{i\phi_{3}}\sin\delta_{1}\sin\delta_{2}\cos\delta_{3}
|10 \rangle +
e^{i\phi_{4}}\sin\delta_{1}\sin\delta_{2}\sin\delta_{3}
|11 \rangle \Big) \; , \nonumber
\end{eqnarray}
where, in Eqs.~(\ref{Phi12}) we have dropped the
subscripts $i,j,k=1,2,3$ $(i\neq j\neq k)$ denoting the three
parties, because in the following we will deal with states invariant
under permutation of any two qubits. In order to evaluate $E_{G}^{(2)}$
for the three-qubit $|W^{(3)}\rangle$ and $|GHZ^{(3)}\rangle$ states
we take advantage of the fact that the coefficients appearing in the
definition of these states are all positive constants. Therefore,
maximization of the overlaps with the states (\ref{Phi12})
does not depend on the phases, that can then be put to zero:
$\Delta=\phi_{q}\,=\,0$ $(q=2,3,4)$. From Eq.~(\ref{overlap2abs}),
we get the following expression of the overlap for the state
$|W^{(3)}\rangle$:
\begin{eqnarray}
\Lambda_{2}^{2}(|W^{(3)}\rangle)=
&&\max_{ \{ \delta_{1}, \delta_{2}, \delta_{3}, \Gamma \} }
\frac{1}{3}\Big[ \cos\delta_{1}
\sin\Gamma + \cos\Gamma \sin\delta_{1} \times \nonumber \\
&& \times (\cos\delta_{2} + \sin\delta_{2} \cos\delta_{3}) \Big]^{2} \, .
\label{Lamb22W3}
\end{eqnarray}
The maximization in Eq.~(\ref{Lamb22W3}) yields the absolute
maximum $\Lambda_{2}^{2}(|W^{(3)}\rangle) = 2/3$. For instance, this
value is reached when
$\delta_{1}=\frac{\pi}{2}$, $\delta_{2}=\frac{\pi}{4}$,
$\delta_{3}=0$, $\Gamma=0$. It is then straightforward to
verify that he three-partite component
of the geometric entanglement present in the three-qubit $W$ state is
\begin{equation}
E_{G}^{(2)}(|W^{(3)}\rangle) = \frac{1}{3} \; .
\label{three-partiteW}
\end{equation}
We see that for three-partite $W$ states the purely three-partite
contribution is strictly lower than the global
geometric entanglement: $E_{G}^{(2)}(|W^{(3)}\rangle)
= 1/3 < E_{G}^{(3)}(|W^{(3)}\rangle) = 5/9$.
On the other hand, for the state $|GHZ^{(3)} \rangle$ the
maximum overlap with the biseparable states is
\begin{eqnarray}
\Lambda_{2}^{2}(|GHZ^{(3)}\rangle)=
&&\max_{ \{ \delta_{1},\delta_{2},\delta_{3}, \Gamma \} }
\frac{1}{2}\Big( \cos\delta_{1} \cos\Gamma  \nonumber \\ && \nonumber \\
&& + \sin\Gamma\sin\delta_{1}\sin\delta_{2}\sin\delta_{3}\Big)^{2} .
\label{Lamb22GHZ3}
\end{eqnarray}
Direct computation yields
\begin{equation}
E_{G}^{(2)}(|GHZ^{(3)}\rangle) = \frac{1}{2} \; .
\label{three-partiteGHZ}
\end{equation}
Thus, in the case of $GHZ$ states we verify that the three-partite
and the global content of geometric entanglement coincide:
$E_{G}^{(2)}(|GHZ^{(3)}\rangle)
= E_{G}^{(3)}(|GHZ^{(3)}\rangle) = 1/2$. This result is an independent
proof that $GHZ$ states possess only genuine tripartite entanglement.
Moreover, we see that the tripartite entanglement of $W$ states
is less than the one of $GHZ$ states: $E_{G}^{(2)}(|W^{(3)}\rangle)
< E_{G}^{(2)}(|GHZ^{(3)}\rangle)$. This result clarifies the original
finding by Wei and Goldbart that the global geometric entanglement
$E_{G}^{(3)}(|W^{(3)}\rangle)$ of $W$ states is larger than the one,
$E_{G}^{(3)}(|GHZ^{(3)}\rangle)$, of $GHZ$ states, and establishes a
proper entanglement hierarchy between the two classes of states.

\begin{figure}[ht]
\centering
\includegraphics*[width=8cm]{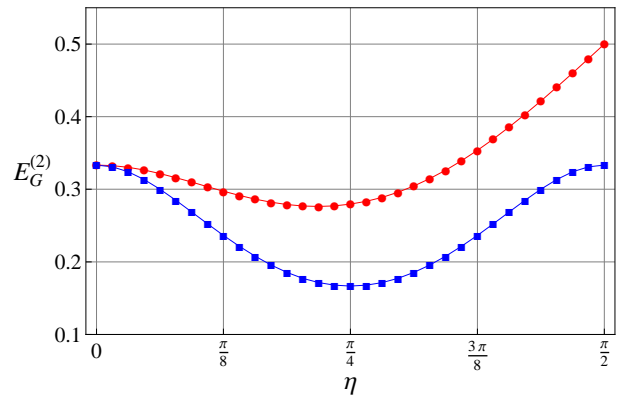}
\caption{(Color online) Behavior of $E_{G}^{(2)}$ for the superpositions
of $|W^{(3)}\rangle$ and $|\tilde{W}^{(3)}\rangle$ (blue line with squares),
Eq.~(\ref{superposWWt}), and for the superpositions of $|W^{(3)}\rangle$ and
$|GHZ^{(3)}\rangle$ (red lines with circles), Eq.~(\ref{superposWGHZ}), as a
function of $\eta$, and for arbitrary phase $\phi$. All plotted quantities are
dimensionless.}
\label{2sepGMESuperpos}
\end{figure}

We now show how the structure of $K$-separability allows to clarify
the nature of the geometric phases in the entanglement of superpositions.
To this aim, let us calculate the distance $E_{G}^{(2)}$ for the
superpositions (\ref{superposWWt}) and (\ref{superposWGHZ}); the
corresponding behavior is reported in Fig. \ref{2sepGMESuperpos} as
a function of $\eta$. Comparing with Fig. \ref{GMESuperposWG}, we
note that both $E_{G}^{(3)}$ and $E_{G}^{(2)}$ exhibit the same
symmetric behavior for the superposition (\ref{superposWWt}), and
acquires a minimum at $\eta=\frac{\pi}{4}$. On the other hand, for the
state (\ref{superposWGHZ}) we observe that $E_{G}^{(2)}$, contrary
to $E_{G}^{(3)}$, is independent of the phase $\phi$.
This implies that the nonlocal nature of the phase $\phi$ is
limited to the set $S_2$ of biseparable states.

\subsection{Symmetric states: $GHZ^{(N)}$ and $W^{(N)}$ states}
\label{subsecB}

In this section we study the properties of the measure (\ref{GMEabs}) for
the states $|GHZ^{(N)}\rangle$ and $|W^{(N)}\rangle$ for arbitrary $N$.
Concerning $GHZ$ states, it is easily verified that, for any $N$,
\begin{equation}
E_{G}^{(K)}(|GHZ^{(N)}\rangle) = \frac{1}{2} \; , \; \; \; K = 2, \ldots, N \; .
\label{GHZEntNqualsiasi}
\end{equation}
Therefore, if we determine the various forms of bipartite and multipartite
entanglement by the geometric measure (\ref{GMEabs}), we obtain that the
$N$-qubit $GHZ$ states possess only $N$-partite entanglement.

Considering $|W^{(N)}\rangle$ states, for a given $N$
all the bipartite and multipartite components of the geometric
entanglement can be evaluated analytically with respect to the
different $K$-separable states. First we study in
detail the $N$-partite entanglement quantified by the distance
$E_{G}^{(2)}(M_1|M_2)$ from the set of biseparable states
$|\Phi \rangle \, = \, |\Phi_{1}^{(M_{1})} \rangle
\otimes |\Phi_{2}^{(M_{2})}\rangle$, for a fixed bipartition
$M_{1}$, $M_{2}=N-M_{1}$,
with $1 \leq M_{1} \leq M_{2} \leq N - 1$.
In this case, using Eq.~(\ref{chi}), $|\Phi\rangle$ takes the
following form
\begin{equation}
|\Phi\rangle \,=\, \sum_{J_{1}=0}^{d_{M_{1}}-1} c_{J_{1}}^{(1)} |J_{1} \rangle\rangle_{M_{1}} \otimes
\sum_{J_{2}=0}^{d_{M_{2}}-1} c_{J_{2}}^{(2)} |J_{2} \rangle\rangle_{M_{2}} \; ,
\label{2sepstM}
\end{equation}
with $c_{J_{s}}^{(s)} =
r_{J_{s}}^{(s)} e^{i\phi_{J_{s}}^{(s)}}$, $s=1,2$, and,
without loss of generality, we let $\phi_{J_{s}}^{(s)} = 0$, $s=1,2$.
By exploiting the decomposition:
\begin{equation}
\sum_{p=0}^{N-1} |2^{p}\rangle\rangle_{N} =
\sum_{p=0}^{M_{2}-1} |0\rangle\rangle_{M_{1}} \otimes |2^{p}\rangle\rangle_{M_{2}}
+ \sum_{p=0}^{M_{1}-1} |2^{p}\rangle\rangle_{M_{1}} \otimes |0\rangle\rangle_{M_{2}}
\nonumber
\end{equation}
one has that the overlap $\Lambda_{2}^{2}(M_1 |M_2)$
can be expressed in the form:
\begin{equation}
\Lambda_{2}^{2}(M_1 |M_2) = \max_{\{ r_{J_{s}}^{(s)} \}}
\frac{1}{N} \Big[ r_{0}^{(1)} \sum_{p = 0}^{M_2 - 1}
r_{2^p}^{(2)} + r_{0}^{(2)} \sum_{p=0}^{M_1 -1} r_{2^p}^{(1)}
\Big]^{2} \; .
\label{2sepruleW}
\end{equation}
The maximization procedure is reported in Appendix \ref{AppendixB},
see Eq.~(\ref{MaxOverlap}).
Using this result, in the case of $2$-separability with respect
to the partitioning $M \otimes (N-M)$, with $M \leq N-M$, the
$K=2$-component of the geometric entanglement in the states
$|W^{(N)}\rangle$ is
\begin{equation}
E_{G}^{(2)}(M |N-M) = \frac{M}{N} \; .
\label{2sepGMEMN-M}
\end{equation}
In the particular instance $M=1$, it immediately follows that
the expression (\ref{2sepGMEMN-M}) realizes the absolute minimum
Eq.~(\ref{GMEabs}), and therefore one has $E_{G}^{(2)}(|W^{(N)}\rangle)
\equiv E_{G}^{(2)}(1 |N-1)= 1/N$, showing that the genuine
$N$-partite geometric entanglement vanishes asymptotically for large
$N$. On the other hand, for the partition obtained by setting
$M = \left[ N/2 \right]$, where $[x]$ denotes the integer part
of $x$, the $K=2$-component of the geometric entanglement tends
to the asymptotic limit $1/2$ for large $N$. The limit coincides with
the maximum possible value, attained by the $|GHZ^{(N)}\rangle$
states.

We turn now to the determination of the generic $K$-components of the
(relative) multipartite geometric entanglement quantified, for arbitrary $K$,
by the distance $E_{G}^{(K)}(M_1|\ldots|M_K)$ from the set of $K$-separable
states for a given partition. We begin by rewriting the generic $K$-separable
state in the form
\begin{equation}
|\Phi \rangle \, = \, \bigotimes_{s=1}^{K} \, \sum_{J_{s}=0}^{d_{M_{s}}-1}
r_{J_{s}}^{(s)} |J_{s}\rangle\rangle_{M_{s}}  \; ,
\label{KsepstM}
\end{equation}
where we remind that the ordering is
$1\leq M_{1} \leq M_{2} \leq \ldots \leq M_{K} \leq N-K+1$,
with $\sum_{s=1}^{K}M_{s}=N$.
In analogy with the previous analysis for the case $K=2$,
it is not difficult to show that the squared overlap
$\Lambda_{K}^{2}(M_1 |\ldots |M_K )$ can be recast in the form
\begin{eqnarray}
&&\Lambda_{K}^{2}(M_1 |\ldots |M_K) = \nonumber \\
&& \nonumber \\
&&\max_{ \{ r_{J_{s}}^{(s)} \} }
\frac{1}{N} \Big[ r_{0}^{(1)}r_{0}^{(2)} \cdots r_{0}^{(K-1)} \;
\sum_{p=0}^{M_{K}- 1} r_{2^{p}}^{(K)} \nonumber \\
&& \nonumber \\
&& + \; r_{0}^{(1)}r_{0}^{(2)} \cdots
\sum_{p=0}^{M_{K-1} -1} r_{2^{p}}^{(K-1)} \; r_{0}^{(K)} + \ldots
\nonumber \\
&& \nonumber \\
&& + \sum_{p=0}^{M_{1} - 1} r_{2^{p}}^{(1)} \; r_{0}^{(2)} \cdots
r_{0}^{(K-1)} r_{0}^{(K)} \Big]^{2} \; .
\label{KsepruleW}
\end{eqnarray}
By a partial maximization, see Appendix \ref{AppendixB}, Eq.~(\ref{KsepruleWbis})
reduces to
\begin{eqnarray}
&&\Lambda_{K}^{2}(M_1 |\ldots |M_K) = \max_{ \{ \delta_{0}^{(s)} \} }
\frac{1}{N} \Big[ \cos\delta_{0}^{(1)} \cos\delta_{0}^{(2)} \cdots \times
\nonumber \\
&& \nonumber \\
&& \times \cos\delta_{0}^{(K-1)} \sin\delta_{0}^{(K)} \sqrt{M_{K}} +
\cos\delta_{0}^{(1)} \cos\delta_{0}^{(2)} \cdots \times
\nonumber \\
&& \nonumber \\
&&\times \cos\delta_{0}^{(K-2)} \sin\delta_{0}^{(K-1)} \cos\delta_{0}^{(K)} \sqrt{M_{K-1}} +\ldots
\nonumber \\
&& \nonumber \\
&&
+ \sin\delta_{0}^{(1)}\cos\delta_{0}^{(2)} \cdots \cos\delta_{0}^{(K-1)} \cos\delta_{0}^{(K)} \sqrt{M_{1}}\Big]^{2} \; .
\label{KsepruleWtris}
\end{eqnarray}
The explicit solution of the problem cannot be given for
generic $K$: One needs to assign a specific value of $K$
in order to solve the problem completely. In principle, full
analytic solutions can always be obtained; however,
the complexity of the problem grows with $K$, so that
for sufficiently large values of $K$ the help of numerical codes
may become necessary. On the other hand, resorting to numerics,
when necessary, poses no particular problem, as all the
equations are rigorously defined and their recursive structures completely
determined. Therefore, the complete analytic and numerical solutions can
always be obtained on demand, for each arbitrarily assigned value of $K$
and $N$, and according to the specific physical problem and type of multipartite
state one is looking at.

Remarkably, from Eq.~(\ref{KsepruleWtris}) it follows that
the multipartite geometric entanglement of  $|W^{(N)}\rangle$ states
satisfies a property of self-similarity and scale invariance.
Namely, given a $N$-qubit $|W^{(N)}\rangle$ state associated to a
partition $M_1 | M_2 |\ldots |M_K $, let us take an integer $L$
and consider the $LN$-qubit state $|W^{(L N)}\rangle$ associated
to the scaled partition $L M_1 | L M_2 |\ldots | L M_K $.
By Eq.~(\ref{KsepruleWtris}), one immediately has that
\begin{equation}
\Lambda_{K}^{2}(M_1 | M_2 |\ldots |M_K ) =
\Lambda_{K}^{2}(L M_1 | L M_2 |\ldots | L M_K) \; .
\label{scaleinvariantoverlap}
\end{equation}
Thus, the $K$-partite geometric measures of entanglement
enjoy the following property of scale invariance:
\begin{equation}
E_{G}^{(K)}(M_1 | M_2 |\ldots |M_K ) =
E_{G}^{(K)}(L M_1 | L M_2 |\ldots | L M_K) \; .
\label{scaleinvariantGME}
\end{equation}
Since relation (\ref{scaleinvariantGME}) applies
for {\it any} partition, it follows that it
holds true for the absolute minimum, Eq.~(\ref{GMEabs}), as well.
Finally, it is worth noticing that the property of scale
invariance of the geometric measures of entanglement is trivially
enjoyed by every $GHZ$ state. \\
Proceeding in the discussion of the general case, we report the
explicit analytic expression for the $K$-component, with $K=3$,
of the multipartite geometric entanglement of $|W^{(N)}\rangle$ states.
The absolute minimum distance $E_{G}^{(3)}(|W^{(N)}\rangle)$ from the set
of all three-separable states ${\mathcal{\mathbf{S}}}_{3}$, that measures the $N$-
and $(N-1)$-partite entanglement of $|W^{(N)}\rangle$ states, reads
\begin{eqnarray}
&& E_{G}^{(3)}(|W^{(N)}\rangle) =  \nonumber \\
&& \nonumber \\
&& \min \big\{ E_{G\,>}^{(3)}(M_1|M_2|M_3),
E_{G\,<}^{(3)}(M_1|M_2|M_3) \big\} \; ,
\label{Trimurti}
\end{eqnarray}
where
\begin{eqnarray}
E_{G\,>}^{(3)}(M_1|M_2|M_3) & = & 1 - \frac{M_{3}}{N} \; , \nonumber \\
M_{3} & \geq & M_{1} + M_{2} \; , \\
&& \nonumber \\
E_{G\,<}^{(3)}(M_1|M_2|M_3) & = &
1 - \frac{4 M_{1} M_{2} M_{3}}{N \Sigma} \; , \nonumber \\
M_{3} & \leq & M_{1} + M_{2} \; ,
\label{EGK32W}
\end{eqnarray}
with $\Sigma = 2(M_{1}M_{2}+M_{1}M_{3}+M_{2}M_{3})-M_{1}^{2}-M_{2}^{2}-M_{3}^{2}$.
The two expressions coincide when $M_{3} = M_1 + M_2$.

In the following we present and discuss the solutions of Eq.~(\ref{KsepruleWtris}).
We will determine the associated $E_{G}^{(K)}(M_1 | M_2 |\ldots | M_K)$ for various
choices of $N$ and $M_1, \ldots, M_K$, and compare them with respect to a
reference standard fixed by the $|GHZ^{(N)}\rangle$ state. Finally, we will
establish for each $N$ the absolute minimum yielding $E_{G}^{(K)}(|W^{(N)}\rangle)$.
\begin{table}[h]
\begin{tabular}{|c|c|c|}
\hline
    & $|GHZ^{(4)}\rangle$ & $|W^{(4)}\rangle$
    \\ \hline
  $E_{G}^{(4)}(1|1|1|1)$  &  $1/2$ & $37/64$ \\ \hline
  $E_{G}^{(3)}(1|1|2)$  &  $1/2$ & $1/2$   \\ \hline
  $E_{G}^{(2)}(2|2)$  &  $1/2$ & $1/2$   \\ \hline
  $E_{G}^{(2)}(1|3)$  &  $1/2$ & $1/4$   \\ \hline
\end{tabular}
\caption{Geometric measures of entanglement $E_{G}^{(K)}(M_1|\ldots|M_K)$,
with $K=2,3,4$, for the $4$-qubit states $|GHZ^{(4)}\rangle$ and $|W^{(4)}\rangle$.}
\label{TabN4}
\end{table}
In Table \ref{TabN4} we report the exact values of the
different geometric entanglements corresponding to
all the possible $K$-partitions in the case $N=4$.
As already stated, in the $|GHZ^{(4)}\rangle$ state the
various components all coincide with the genuine $4$-partite entanglement.
In the $|W^{(4)}\rangle$ state one has that for $K=3$ and $K=4$, due to the symmetry
under exchange of any pair of qubits, there is a unique way to partition the
system, and the relative component of the geometric entanglement coincides with
the absolute component. In the case $K=2$ one has two inequivalent possible
partitions, and, as already proved in the general case, the absolute minimum is attained
for the partition $1|3 \equiv 1|(N-1)$.

In Tables \ref{TabN5} and \ref{TabN6} we
report the different multipartite components of the geometric entanglement,
respectively in the $|W^{(5)}\rangle$ and $|W^{(6)}\rangle$ states. The fixed
reference value $1/2$ of the $|GHZ\rangle$ states is not reported.
\begin{table}[h]
\begin{tabular}{|c|c||c|c|}
\hline
    & $|W^{(5)}\rangle$ &  & $|W^{(5)}\rangle$
    \\ \hline
  $E_{G}^{(5)}(1|1|1|1|1)$  &  $0.590$ & $E_{G}^{(3)}(1|1|3)$ & $2/5$ \\ \hline
  $E_{G}^{(4)}(1|1|1|2)$  &  $0.559$ & $E_{G}^{(2)}(2|3)$   & $2/5$ \\ \hline
  $E_{G}^{(3)}(1|2|2)$  &  $19/35$ &  $E_{G}^{(2)}(1|4)$  & $1/5$ \\ \hline
  \end{tabular}
\caption{Geometric measures of entanglement $E_{G}^{(K)}(M_1|\ldots|M_K)$,
with $K=2,3,4,5$, for the state $|W^{(5)}\rangle$.}
\label{TabN5}
\end{table}
From Tables \ref{TabN5} and \ref{TabN6} we see that for $N\geq5$ there appear
sets $S_K(M_1|\ldots|M_K)$ containing inequivalent partitions also for
$K>2$. Moreover, we observe that the relative distances do not obey a definite
hierarchy; for instance, from Table \ref{TabN6} we see that
$E_{G}^{(3)}(2|2|2) > E_{G}^{(4)}(1|1|1|3)$. However, and more importantly,
the hierarchy of absolute distances is never violated. For instance,
$\min E_{G}^{(3)}(M_1|M_2|M_3) < \min E_{G}^{(4)}(M_1|M_2|M_3|M_4)$, in perfect
agreement with the general ordering established by Eq.~(\ref{MatchPoint}).
\begin{table}[h]
\begin{tabular}{|c|c||c|c|}
\hline
    & $|W^{(6)}\rangle$ & & $|W^{(6)}\rangle$
    \\ \hline
  $E_{G}^{(6)}(1|1|1|1|1|1)$  &  $0.598$ & $E_{G}^{(3)}(1|2|3)$ & $1/2$ \\ \hline
  $E_{G}^{(5)}(1|1|1|1|2)$  &  $0.580$ & $E_{G}^{(2)}(3|3)$  & $1/2$ \\ \hline
  $E_{G}^{(4)}(1|1|2|2)$  & $0.567$  & $E_{G}^{(3)}(1|1|4)$  &  $1/3$  \\ \hline
  $E_{G}^{(3)}(2|2|2)$   & $5/9$  &  $E_{G}^{(2)}(2|4)$  & $1/3$ \\ \hline
  $E_{G}^{(4)}(1|1|1|3)$ &  $1/2$&  $E_{G}^{(2)}(1|5)$  & $1/6$  \\ \hline
  \end{tabular}
\caption{Geometric measures of entanglement $E_{G}^{(K)}(M_1|\ldots|M_K)$, with
$K=2,3,4,5,6$, for the state $|W^{(6)}\rangle$.}
\label{TabN6}
\end{table}
Finally, we remark that all the measures
evaluated analytically are rational numbers, and that the ones computed numerically
appear to be approximations of rational numbers. Therefore, we conjecture that, for
every {\it finite} $N$, all the relative and absolute multipartite geometric measures
of entanglement are expressed by rational numbers.

\subsection{Superpositions of $W^{(N)}$ and $GHZ^{(N)}$ states}
\label{subsecGHZW}

\begin{figure}[h]
\centering
\includegraphics*[width=8cm]{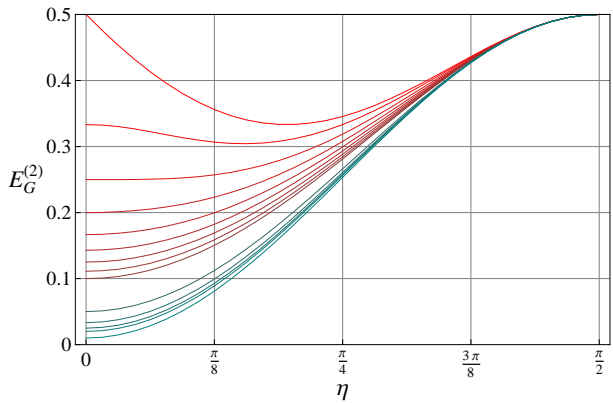}
\caption{(Color online) Behavior of $E_{G}^{(2)}(1|N-1)$ in the superposition
states Eq.~(\ref{supWGHZ}), as a
function of $\eta$, for $N=2,3,\ldots,10$, and for
$N = 20,30,40,50,100$. Curves are ordered from top to bottom with increasing
$N$, with the uppermost curve corresponding to $N=2$ and the lowermost curve
corresponding to $N=100$. All plotted quantities are dimensionless.}
\label{2sepGMESuperposN}
\end{figure}

It is of interest to investigate symmetric states constituted
by generic superpositions of $W$ and $GHZ$ states:
\begin{equation}
|WGHZ^{(N)}\rangle \,=\, \cos\eta |W^{(N)}\rangle + \sin\eta |GHZ^{(N)}\rangle \,,
\label{supWGHZ}
\end{equation}
where the $N$-qubit $GHZ^{(N)}$ and $W^{(N)}$ states are defined
by Eqs.~(\ref{GHZNbin}) and (\ref{WNbin}), respectively.
The squared overlap $\Lambda_{2}^{2}(M_{1}|M_{2})$ associated
with the set of bi-separable states (\ref{2sepstM}) can be
computed exactly. One has:
\begin{eqnarray}
&&\Lambda_{2}^{2}(M_{1}|M_{2}) = \max_{\{ r_{J_{s}}^{(s)} \}}
\Big[\frac{\cos\eta}{\sqrt{N}} \Big( r_{0}^{(1)} \sum_{p = 0}^{M_2 - 1}
r_{2^p}^{(2)} + \label{overlapWGHZN} \\
&& \nonumber \\
&&r_{0}^{(2)} \sum_{p=0}^{M_1 -1} r_{2^p}^{(1)} \Big) +
\frac{\sin\eta}{\sqrt{2}} \Big(r_{0}^{(1)}r_{0}^{(2)} +  r_{2^{M_{1}} -1 }^{(1)} r_{2^{M_{2}} -1 }^{(2)} \Big)
\Big]^{2} \; . \nonumber
\end{eqnarray}
As shown in Appendix \ref{AppendixB}, a partial maximization procedure
reduces the above relation to:
\begin{eqnarray}
&&\Lambda_{2}^{2}(M_{1}|M_{2}) = \max_{\{ \delta_{q}^{(s)} \}}
\Big|\frac{\cos\eta}{\sqrt{N}} \Big(\cos\delta_{0}^{(1)}\sin\delta_{0}^{(2)}\sin\delta_{1}^{(2)} \times
\nonumber \\
&& \nonumber \\
&& \sqrt{M_{2}} + \sin\delta_{0}^{(1)}\sin\delta_{1}^{(1)}\sin\delta_{0}^{(2)} \sqrt{M_{1}} \Big)
+\frac{\sin\eta}{\sqrt{2}}\Big( \cos\delta_{0}^{(1)}  \times \nonumber  \\
&& \nonumber \\
&& \times \cos\delta_{0}^{(2)}+ \sin\delta_{0}^{(1)}\cos\delta_{1}^{(1)}
\sin\delta_{0}^{(2)}\cos\delta_{1}^{(2)} \Big)
\Big|^{2} \,.
\label{overlapWGHZN2}
\end{eqnarray}
It is rather straightforward to prove that Eq.~(\ref{overlapWGHZN2}) and
the associated geometric measure of entanglement enjoy a property of scale
invariance: $\Lambda_{2}^{2}(M_{1}|M_{2})=\Lambda_{2}^{2}(L M_{1}| L M_{2})$
(with $L$ integer), where $\Lambda_{2}^{2}(L M_{1}| L M_{2})$ is the squared
overlap associated to the $LN$-qubit superposition state $|WGHZ^{(LN)}\rangle$.
At fixed values of $\eta$, $M_{1}$, and $M_{2}$, numerical evaluation of
Eq.~(\ref{overlapWGHZN2}) can always be carried out easily.
Being particularly interested in the quantification of genuine multipartite
entanglement, we evaluate the geometric measure $E_{G}^{(2)}(1|N-1)$, and
report it in Fig.~\ref{2sepGMESuperposN} for different values of $N$.
For $N=2$, the state $|WGHZ^{(2)}\rangle$ reduces to the superposition of Bell states.
In this case, $E_{G}^{(2)}(1|1)$ attains the maximum value $1/2$ at $\eta=0,\frac{\pi}{2}$.
For $N=3$, the curve coincides with the one plotted in Fig.~\ref{2sepGMESuperpos}.
Let us notice that in the instances $N=2,3$, absolute minima exist in the interval $(0,\frac{\pi}{2})$.
For $N\geq 4$, $E_{G}^{(2)}(1|N-1)$ increases monotonically from the value $1/N$
attained in the $|W^{(N)}\rangle$ state to the value $1/2$ attained in
the $|GHZ^{(N)}\rangle$ state.

\subsection{$N = 4$ cluster state}
\label{subsecC}

In this subsection we apply the formalism previously introduced to
the determination of the multipartite geometric entanglement of
$N$-qubit cluster states \cite{BrieRaus} in the case $N=4$, the
only nontrivial instance that allows an explicit closed expression.
In fact, the $N=4$ cluster state can be expressed as a
superposition of the form
\begin{equation}
|Cls^{(4)}\rangle = \frac{1}{2} (|0000\rangle + |0011\rangle +
|1100\rangle - |1111\rangle) \; .
\label{Cls4}
\end{equation}
Recently, this state has been produced and characterized
experimentally \cite{4qubitClusterGen,Report}, as a relevant
representative of the class of stabilizer states, which are very important
both from a theoretical perspective and from a practical point of view
for their property of entanglement persistency and for the implementation
of one-way quantum computation \cite{1wayQC}.

In Table \ref{TabCluster} we report the values of the different
components of the geometric entanglement in the $N=4$ cluster state
corresponding to all the possible $K$-partitions of the $4$-partite system.
\begin{table}[h]
\begin{tabular}{|c|c|c|c|c|}
\hline
    & $E_{G}^{(4)}(1|1|1|1)$ & $E_{G}^{(3)}(1|1|2)$ & $E_{G}^{(2)}(2|2)$ &  $E_{G}^{(2)}(1|3)$
    \\ \hline
 $|Cls^{(4)}\rangle$  & $3/4$  & $1/2$ & $1/2$  & $1/2$ \\ \hline
\end{tabular}
\caption{Multipartite geometric measures of entanglement $E_{G}^{(K)}(M_1|\ldots|M_K)$,
for $K=2,3,4$, in the $4$-qubit cluster state $|Cls^{(4)}\rangle$.}
\label{TabCluster}
\end{table}
We observe that in the case of $N=4$ cluster states there is a degeneracy
in the geometric structure, as the absolute minimum is realized not only
by the genuine four-partite components of entanglement $E_{G}^{(2)}(1|3)$
and $E_{G}^{(2)}(2|2)$, but also by the three-partite
component $E_{G}^{(3)}(|Cls^{(4)}\rangle)$. The latter coincides with
$E_{G}^{(2)}(|Cls^{(4)}\rangle)$. On the other hand, as
$E_{G}^{(4)}(|Cls^{(4)}\rangle) > E_{G}^{(3)}(|Cls^{(4)}\rangle)$, the
$4$-qubit cluster state possesses also a bipartite component besides
the genuine four-partite contribution.

\subsection{Magnon states}
\label{magnoni}

Going further towards higher generalizations that are physically significant,
we discuss the class of symmetric $N$-qubit entangled states expressed as superpositions
of magnon states \cite{Mattis,ManyMagnons}. A magnon is an elementary excitation of
magnetic materials, i.e. a quantum of a spin wave, and $W$ states are actually
the simplest superpositions of all possible magnon states containing only one excitation.
In the generic case of $k$ excitations on $N$ particles, the multi-magnon superposition
states can be written in the form:
\begin{eqnarray}
 {}\hspace{-1cm}|Mg_{k}^{(N)}\rangle &= & \Big( \begin{array}{c}
                             N \\
                             k
                           \end{array} \Big)^{-1/2}
                           \sum_{p_{k} = k - 1}^{N-1} \quad \sum_{p_{k-1} = k - 2}^{p_{k} - 1}
                           \cdots \nonumber \\
                           && \cdots \sum_{p_{1} = 0}^{p_{2} -1} \,
                           |2^{p_{k}} + 2^{p_{k-1}} + \ldots + 2^{p_{1}} \rangle \rangle_{N} \,,
\label{kmagnonstate}
\end{eqnarray}
For the sake of illustration, let us consider explicitly the case $k=2$, i.e. the
superpositions of all possible $N$-qubit states containing two elementary excitations.
Such states can be expressed in the form:
\begin{equation}
|Mg_{2}^{(N)}\rangle \,=\, \Big( \begin{array}{c}
                             N \\
                             2
                           \end{array} \Big)^{-1/2}
                           \sum_{p=1}^{N-1}\sum_{q=0}^{p-1} \, |2^{p}+2^{q}\rangle\rangle_{N} \,,
\label{2magnonstate}
\end{equation}
with $N\geq 4$. \\
In Table \ref{TabMagn4} we report the values, computed numerically, of the
different components of the geometric entanglement in the $4$-qubit
two-magnon state $|Mg_{2}^{(4)}\rangle$.
\begin{table}[h]
\begin{tabular}{|c|c|c|c|c|}
\hline
    & $E_{G}^{(4)}(1|1|1|1)$ & $E_{G}^{(3)}(1|1|2)$ & $E_{G}^{(2)}(2|2)$ &  $E_{G}^{(2)}(1|3)$
    \\ \hline
 $|Mg_{2}^{(4)}\rangle$  & $0.625$  & $0.583$ & $1/3$  & $1/2$ \\ \hline
\end{tabular}
\caption{Geometric measures of entanglement $E_{G}^{(K)}(M_1|\ldots|M_K)$, with
$K=2,3,4$, for the $4$-qubit magnon state $|Mg_{2}^{(4)}\rangle$.}
\label{TabMagn4}
\end{table}
By comparing Tables \ref{TabMagn4}, \ref{TabN4}, and \ref{TabCluster},
we see that, with respect to the $|W^{(4)}\rangle$ state, the $|Mg_{2}^{(4)}\rangle$ state
possesses enhanced genuine multipartite entanglement:
\begin{eqnarray}
E_{G}^{(2)}(|Mg_{2}^{(4)}\rangle) & = & \min\{E_{G}^{(2)}(2|2)\, ; \, E_{G}^{(2)}(1|3)\} \nonumber \\
& = & 1/3 > E_{G}^{(2)}(|W^{(4)}\rangle) = 1/4 \, . \nonumber
\end{eqnarray}
Analogously, the three-partite component $E_{G}^{(3)}(1|1|2)$ is enhanced in the magnon
state compared to the $W$ state. On the contrary, the $|Mg_{2}^{(4)}\rangle$ state
possesses a smaller amount of genuine multipartite entanglement
compared to the $|Cls^{(4)}\rangle$ state,
while the three-partite component $E_{G}^{(3)}(1|1|2)$ is larger. \\
Next, we generalize the previous analysis to the case of $N$-qubit two-magnon states
(\ref{2magnonstate}) with arbitrary $N$, and determine the
geometric measure of entanglement $E_{G}^{(2)}(M_{1}|M_{2})$,
i.e. the distance from the set of $2$-separable states of the form (\ref{2sepstM}).
By exploiting the decomposition:
\begin{eqnarray}
&&\sum_{p=1}^{N-1}\sum_{q=0}^{p-1} |2^{p}+2^{q}\rangle\rangle_{N} =
\sum_{p=1}^{M_{2}-1}\sum_{q=0}^{p-1} |0\rangle\rangle_{M_{1}} \otimes |2^{p}+2^{q}\rangle\rangle_{M_{2}}
\nonumber \\
&& \nonumber \\
&&+ \sum_{p=1}^{M_{1}-1}\sum_{q=0}^{p-1} |2^{p}+2^{q}\rangle\rangle_{M_{1}} \otimes |0\rangle\rangle_{M_{2}}
\nonumber \\
&& \nonumber \\
&& +\sum_{p=0}^{M_{1}-1}|2^{p}\rangle\rangle_{M_{1}} \otimes  \sum_{p=0}^{M_{2}-1}|2^{p}\rangle\rangle_{M_{2}} \;,
\end{eqnarray}
the squared overlap $\Lambda_{2}^{2}(M_{1}|M_{2})$ writes:
\begin{eqnarray}
&&\Lambda_{2}^{2}(M_{1}|M_{2}) \,=\, \left(\begin{array}{c}
                                       N \\
                                       2
                                     \end{array}\right)^{-1}
\max_{\{ r_{J_{s}}^{(s)} \}}\left[r_{0}^{(1)} \sum_{p=1}^{M_{2}-1} \sum_{q=0}^{p-1} r_{2^{p}+2^{q}}^{(2)} \right.   \nonumber \\
&& \nonumber \\
&&\left. + r_{0}^{(2)} \sum_{p=1}^{M_{1}-1} \sum_{q=0}^{p-1} r_{2^{p}+2^{q}}^{(1)}
+\sum_{p=0}^{M_{1}-1} r_{2^{p}}^{(1)} \sum_{p=0}^{M_{2}-1} r_{2^{p}}^{(2)} \right]^{2} .
\label{overlap2sepMagnon}
\end{eqnarray}
The mathematical details concerning the maximization of Eq.~(\ref{overlap2sepMagnon})
are treated in Appendix \ref{AppendixMagnon}.
From Eq.~(\ref{overlap2sepMagnon5}), fixing $M=M_{1}\leq M_{2}=N-M$, we obtain:
\begin{eqnarray}
&&\Lambda_{2}^{2}(M|N-M) \,=\,
[N(N-1)]^{-1} \nonumber \\
&& \nonumber \\
&&\max \{ (N-M)(N-M-1) \, ; \,  2M(N-M) \}  \,.
\label{overlap2sepMagnFinale}
\end{eqnarray}
As in the previous instances of symmetric states,
the squared overlap $\Lambda_{2}^{2}(M_{1}|N_{2})$
and the corresponding geometric measure satisfy the property of scale invariance
also for magnon states. One has:
$\Lambda_{2}^{2}(M_{1}|N_{2})=\Lambda_{2}^{2}(L M_{1}|L N_{2})$ (with $L$ integer),
$\Lambda_{2}^{2}(L M_{1}|L N_{2})$ being the squared overlap associated with the
$LN$-qubit two-magnon state $|Mg_{2}^{(LN)}\rangle$.
For $M_{1}=1$ and $M_{2}=N-1$,
the relation (\ref{overlap2sepMagnFinale}) reduces to
$E_{G}^{(2)}(1|N-1)= 2/N$, with $N\geq 4$. Therefore,
the genuine $N$-partite geometric entanglement contained in two-magnon
states vanishes asymptotically in the limit of large $N$, analogously to
the case of $N$-qubit $W$ states. This property is expected to hold
in every multi-magnon state: At any fixed, finite value of $k$, the
genuine $N$-partite entanglement contained in a $k$-magnon state vanishes
in the limit of large $N$.

\section{Asymmetric states: generalized $W$-like superposition states}
\label{SecGeneralW}

In this Section, we evaluate the geometric measure of entanglement for
the class of asymmetric, generalized $N$-qubit $W$-like superposition
states defined as
\begin{equation}
|\psi_{W}^{(N)}\rangle \,=\, \mathcal{N}_{N} \, \sum_{p=0}^{N-1} \, \gamma_{p+1} e^{i \xi_{p+1}}\,
|2^{p}\rangle\rangle_{N} \,,
\label{WNlikedef}
\end{equation}
where $\gamma_{p}$ are real parameters, $\xi_{p}$ are real phases,
and the normalization factor is
$\mathcal{N}_{N} = \big(\sum_{p=0}^{N-1} \, \gamma_{p+1}^{2}\big)^{-1/2}$.
These states play a relevant role in quantum information science according to
the following considerations. It is well known that true tripartite entanglement
of the state of a system of three qubits can be classified on the basis of stochastic
local operations and classical communications. Such states can then be classified into
two categories corresponding to the $GHZ$ and $W$ states. It is known that $GHZ$ states
can be used for teleportation and superdense coding, but the standard symmetric $W$ states
cannot. However, it has been shown that the class of asymmetric, generalized $W$-like
superposition states (\ref{WNlikedef}) can be used as entangled resources for the
implementation of perfect teleportation and superdense coding \cite{GeneralWstates}.
Moreover, several methods for their preparation have been proposed \cite{WstatePrepar}.

\begin{figure}[ht]
\centering
\includegraphics*[width=8cm]{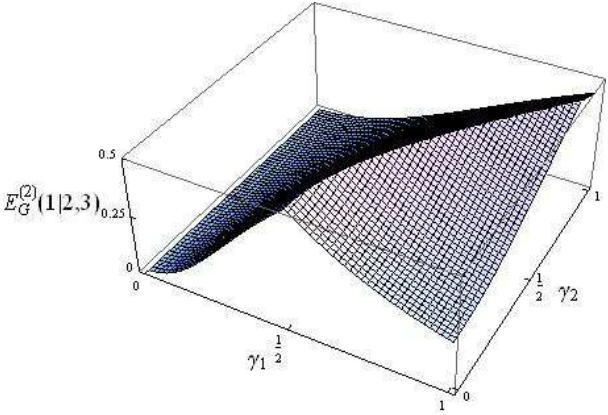}
\caption{(Color online) The relative measure of geometric entanglement
$E_{G}^{(2)}(1|2,3)$ for the state $|\psi_{W}^{(3)}\rangle$, plotted
as a function of $\gamma_{1}$ and $\gamma_{2}$,
at fixed $\gamma_{3}=1/2$. All plotted quantities are dimensionless.}
\label{Fig3WlikeGME1_23}
\end{figure}

\begin{figure}[h]
\centering
\includegraphics*[width=8cm]{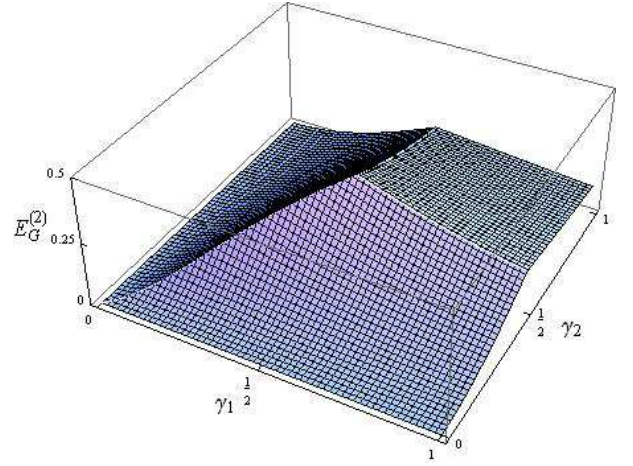}
\caption{(Color online) The absolute measure of geometric entanglement
$E_{G}^{(2)} (|\psi_{W}^{(3)}\rangle)$, plotted
as a function of $\gamma_{1}$ and $\gamma_{2}$,
at fixed $\gamma_{3}=1/2$. All plotted quantities are dimensionless.}
\label{Fig3WlikeEG2}
\end{figure}

\begin{figure}[h]
\centering
\includegraphics*[width=8cm]{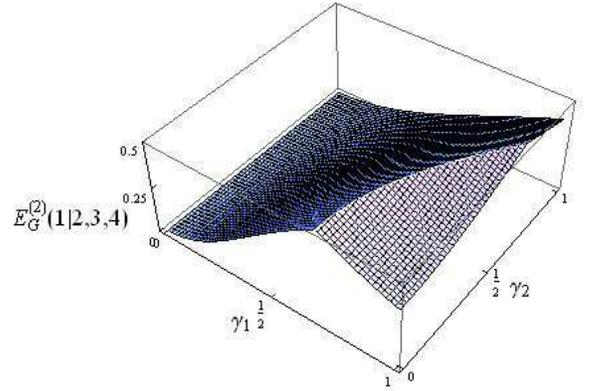}
\caption{(Color online) The relative measure of geometric entanglement
$E_{G}^{(2)}(1|2,3,4)$ for the state $|\psi_{W}^{(4)}\rangle$, plotted
as a function of $\gamma_{1}$ and $\gamma_{2}$,
at fixed $\gamma_{3} =2/3$ and $\gamma_{4} = 1/6$. All plotted quantities
are dimensionless.}
\label{Fig4WlikeGME1_234}
\end{figure}

\begin{figure}[h]
\centering
\includegraphics*[width=8cm]{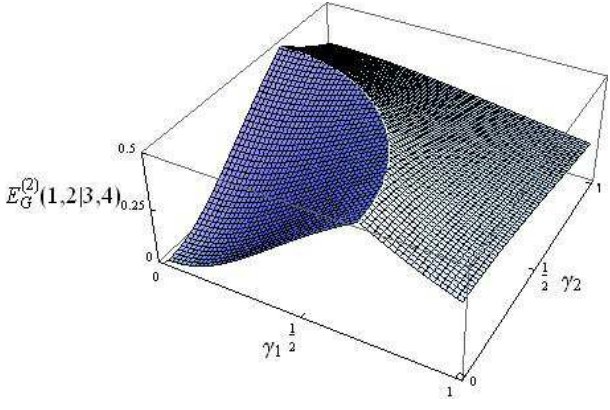}
\caption{(Color online) The relative measure of geometric entanglement
$E_{G}^{(2)}(1,2|3,4)$ for the state $|\psi_{W}^{(4)}\rangle$, plotted
as a function of $\gamma_{1}$ and $\gamma_{2}$,
at fixed $\gamma_{3} =2/3$ and $\gamma_{4} = 1/6$.
All plotted quantities are dimensionless.}
\label{Fig4WlikeGME12_34}
\end{figure}

Without loss of generality and information content in the definition of the state,
we assume $\gamma_{p} \in [0,1]$ and $\xi_{p} \in [0,2\pi]$.
Moreover, in the following, we will let $\xi_{p}=0$ as it can be shown that the phases
are irrelevant in the calculation of the geometric measures,
being always canceled by the free phases of the $K$-separable states
in the maximization procedure.
The states (\ref{WNlikedef}) are asymmetric, i.e. not invariant with respect
the permutation of any couple of qubit.
We first give explicit examples of application for
the three-qubit and four-qubit instances, Eq.~(\ref{WNlikedef}) with $N=3,4$ respectively.
In the three-qubit case, proceeding as in subsection \ref{subsecA},
we compute the squared overlap (\ref{overlap2rel}) for the state $|\psi_{W}^{(3)}\rangle$.
Dealing with asymmetric states, we have to specify the elementary qubits
contained in the two sets $Q_{1}$ and $Q_{2}$ which determine the set $S_{2}(Q_{1}|Q_{2})$
of the $2$-separable states. Thus, we compute the quantity
$\Lambda_{2}^{2}(i|j,k)$, where $i,j,k=1,2,3$, with $i\neq j \neq k$ denote the three
elementary qubits.
The calculation of this quantity yields:
\begin{equation}
\Lambda_{2}^{2}(i|j,k) \,=\, \mathcal{N}_{3}^2
\max \{\gamma_{i}^{2} \,, \gamma_{j}^{2}+\gamma_{k}^{2}\} \,.
\label{overlapW3like}
\end{equation}
In Fig.~\ref{Fig3WlikeGME1_23}, we plot the relative geometric measure $E_{G}^{(2)}(1|2,3)$
for the state $|\psi_{W}^{(3)}\rangle$ as a function of the variables
$\gamma_{1}$ and $\gamma_{2}$, at a fixed value of $\gamma_{3}$.
We see that $E_{G}^{(2)}(1|2,3)$ is formed by two surfaces whose curve of separation
stays at the maximum attainable value $1/2$.
Similar plots can be obtained for $E_{G}^{(2)}(2|1,3)$ and $E_{G}^{(2)}(3|1,2)$.
The absolute geometric measure of entanglement, as defined in Eq.~(\ref{GMEabs}),
is given by $E_{G}^{(2)} (|\psi_{W}^{(3)}\rangle) \,=\, \min_{ \{i,j,k\} } \{ E_{G}^{(2)}(i|j,k) \}$ with $i,j,k=1,2,3$ and $i\neq j\neq k$; the absolute minimum is evaluated with respect
to all possible permutations of the indices, as the state is not symmetric under
the exchange of any two qubits. The absolute measure of geometric entanglement
with respect to biseparable states is plotted in Fig.~\ref{Fig3WlikeEG2}.
The surface describing $E_{G}^{(2)} (|\psi_{W}^{(3)}\rangle)$ is formed by
the contributions of three surfaces whose common intersection
is at the absolute maximum $1/3$.
Let us notice that this absolute maximum is always achieved
for $\gamma_{1}=\gamma_{2}=\gamma_{3}$.
In the case $N=4$, besides the bipartition $1|3$,
we also have to take into account the bipartition $2|2$.
Direct evaluation yields:
\begin{eqnarray}
&&\Lambda_{2}^{2}(i|j,k,l) \,=\, \mathcal{N}_{4}^2
\max \{\gamma_{i}^{2} \,, \gamma_{j}^{2}+\gamma_{k}^{2}+\gamma_{l}^{2}\} \,,
\label{overlapW4like13} \\
&& \nonumber \\
&&\Lambda_{2}^{2}(i,j|k,l) \,=\, \mathcal{N}_{4}^2
\max \{\gamma_{i}^{2}+\gamma_{j}^{2} \,, \gamma_{k}^{2}+\gamma_{l}^{2}\} \,.
\label{overlapW4like22}
\end{eqnarray}
The relative geometric measures $E_{G}^{(2)}(1|2,3,4)$ and $E_{G}^{(2)}(1,2|3,4)$
are plotted in Figs.~\ref{Fig4WlikeGME1_234} and \ref{Fig4WlikeGME12_34}, respectively,
as functions of $\gamma_{1}$ and $\gamma_{2}$, at fixed $\gamma_{3}$ and $\gamma_{4}$.
Similarly to the plot in Fig.~\ref{Fig3WlikeGME1_23},
these three-dimensional plots are characterized by
two surfaces whose separation curve is set at the maximum value $1/2$.
The absolute geometric measure
$E_{G}^{(2)} (|\psi_{W}^{(4)}\rangle)$ exhibits a behavior similar to that observed
for the three-qubit instance, see Fig.~\ref{Fig3WlikeEG2}.
We finally consider the general instance of $N$-qubit states,
expressed by Eq.~(\ref{WNlikedef}) for arbitrary $N$ (with $\xi_{p+1}=0$).
We compute the squared overlap $\Lambda_{K}^{2}(Q_1 |\ldots |Q_K )$
associated with the $K$-separable state (\ref{KseparablePhi}).
For simplicity, we choose $Q_{1} = \{1,\ldots, M_{1}\}$, $Q_{2} = \{M_{1}+1,\ldots, M_{1}+M_{2}\}$,
$\ldots$, $Q_{K} = \{M_{1}+\ldots +M_{K-1}+1,\ldots, N\}$.
Direct evaluation yields:
\begin{eqnarray}
&&\Lambda_{K}^{2}(Q_1 |\ldots |Q_K) = \nonumber \\
&& \nonumber \\
&&\max_{ \{ r_{J_{s}}^{(s)} \} }
\mathcal{N}_{N}^2 \Big[ r_{0}^{(1)}r_{0}^{(2)} \cdots r_{0}^{(K-1)} \;
\sum_{p=0}^{M_{K}- 1} r_{2^{p}}^{(K)} \gamma_{p+1} \nonumber \\
&& \nonumber \\
&& + \; r_{0}^{(1)}r_{0}^{(2)} \cdots
\sum_{p=0}^{M_{K-1} -1} r_{2^{p}}^{(K-1)}\gamma_{p+1+M_{K}} \; r_{0}^{(K)} + \ldots
\nonumber \\
&& \nonumber \\
&& + \sum_{p=0}^{M_{1} - 1} r_{2^{p}}^{(1)} \gamma_{p+1+M_{2}+\ldots +M_{K}} \; r_{0}^{(2)} \cdots
 r_{0}^{(K)} \Big]^{2} .
\label{KsepruleWlike}
\end{eqnarray}
The partial maximization of Eq.~(\ref{KsepruleWlike}), see Appendix \ref{AppendixB},
yields the relation
\begin{eqnarray}
&&\Lambda_{K}^{2}(Q_1 |\ldots |Q_K) = \max_{ \{ \delta_{0}^{(s)} \} }
\mathcal{N}_{N}^2 \left[ \cos\delta_{0}^{(1)}  \cdots \cos\delta_{0}^{(K-1)}\times \right.
\nonumber \\
&& \nonumber \\
&& \times  \sin\delta_{0}^{(K)} \sqrt{\sum_{p=0}^{M_{K}-1}\gamma_{p+1}^{2}} +
\cos\delta_{0}^{(1)}  \cdots \cos\delta_{0}^{(K-2)}\times
\nonumber \\
&& \nonumber \\
&&\times  \sin\delta_{0}^{(K-1)} \cos\delta_{0}^{(K)} \sqrt{\sum_{p=0}^{M_{K-1}-1}\gamma_{p+1+M_{K}}^{2}} +\ldots
\nonumber \\
&& \nonumber \\
&& \left.
+ \sin\delta_{0}^{(1)}\cos\delta_{0}^{(2)} \cdots  \cos\delta_{0}^{(K)} \sqrt{\sum_{p=0}^{M_{1}-1}\gamma_{p+1+M_{2}+\ldots+M_{K}}^{2}} \, \right]^{2}  .
\nonumber \\
\label{KsepruleWlike2}
\end{eqnarray}
As with Eq.~(\ref{KsepruleWtris}),
in order to obtain the final result from the simplified relation (\ref{KsepruleWlike2}),
one needs to perform a (numerical) maximization only over $K$ variables.
It is straightforward to observe that Eq.~(\ref{KsepruleWlike2}) reduces
to Eq.~(\ref{KsepruleWtris}) for $\gamma_{p+1}=1$ for every $p$.
Moreover, given Eq.~(\ref{KsepruleWlike2}), results (\ref{overlapW3like}),
(\ref{overlapW4like13}), and (\ref{overlapW4like22}) are immediately
recovered as particular cases.

\section{Conclusions and Outlook}

\label{Conclusions}

In this work we have introduced and discussed a class of generalized
geometric measures of entanglement. For pure quantum states of $N$
elementary subsystems, these extended measures are defined as the
distances from the sets of $K$-separable states ($K=2,\ldots,N$).
In principle, the entire set of these $N-1$ geometric measures provides
a complete quantification and a hierarchical ordering of
the different bipartite and multipartite components of the global
geometric entanglement, and allows to discriminate among the
different multi-party contributions. After introducing and elucidating
the fundamental properties of the generalized geometric measures, we
have investigated in detail multipartite pure states of $N$-qubit
systems. For the multi-qubit case, we have derived some general properties
of the extended geometric measures and discussed a systematic method for their
evaluation in symmetric states including $W(N)$ states, $GHZ(N)$ states,
and their superpositions; symmetric cluster states; and multi-magnon states.
Moreover, considering asymmetric states, we have introduced a method for the
systematic determination of the multipartite components of geometric entanglement
in a large class of generalized $W$-like superposition states.
We have identified a property of self-similarity and scale-invariance holding
for all types of geometric entanglement and symmetric multipartite pure states
of many-qubit systems. Finally, in a series of mathematical appendices, we have
sketched the main mathematical framework needed for the exact numerical and/or
analytical evaluation of the bipartite and multipartite components of geometric
entanglement in any, arbitrarily chosen, pure state of many-qubit systems.

A challenge of great potential interest is to extend the mathematical framework
and the recursive computation schemes developed for multi-qubit states
to higher-dimensional quantum systems. A further significant issue concerns
the extension of the geometric setting to mixed states, beyond the immediate,
but in practice not very useful, procedures based on the convex hull
construction. Devising alternative, but conceptually equally satisfactory,
extensions would be especially important in order to establish a deeper
understanding of the possible operational characterizations for the set
of geometric measures of entanglement in the presence of classical noise and
non unitary quantum operations. Finally, it would be important to investigate
to what extent the novel multi-component measures of geometric entanglement
defined in the present work could be exploited to construct geometric
monotones obeying a structure of shared entanglement and monogamy bounds for
distributed entanglement, and their role in the understanding of quantum
critical phenomena and quantum cooperative systems.

\appendix

\section{Hyperspherical parametrization}
\label{AppHypPar}

A very convenient parametrization for the moduli $r_{J}$ in Eq.~(\ref{chi}),
that automatically solves the normalization constraint,
is the representation in hyperspherical coordinates in $d_{M}$ dimensions:
\begin{eqnarray}
v_{0} &=& \cos\delta_{0} \,, \nonumber \\
v_{1} &=& \sin\delta_{0}\cos\delta_{1} \,, \nonumber \\
&& \ldots \,, \nonumber \\
v_{d_{M} - 2} &=& \sin\delta_{0}\cdots \sin\delta_{d_{M} - 3}\cos \delta_{d_{M} - 2}
\,, \nonumber \\
v_{d_{M}-1} &=& \sin\delta_{0}\cdots \sin\delta_{d_{M} - 3}\sin\delta_{d_{M} - 2}
\,, \label{hypersphparam}
\end{eqnarray}
where $\delta_{L}$ are angles with values in the interval
$[0,\frac{\pi}{2}]$.
Let us notice that the one-to-one mapping
between $r_{J}$ and $v_{L}$ (with $J,L=0,\ldots,d_{M}-1$),
or equivalently between the indices $J$ and $L$,
can be chosen by following a suitable ordering.
In fact, the term $v_{L}$ will be constituted by a product of $L$ trigonometric functions;
thus, it would be convenient to have expressions involving parameters $v_{L}$
with low values of the index $L$.
Considering the state (\ref{chi}), the most immediate choice for $r_{J}$
is obtained by letting $r_{J}=v_{J}$ for $J=0,\ldots,d_{M}-1$. \\
Interpreting the two levels $0$ and $1$ of the single qubit
as the ground and the excited levels, respectively, of the elementary system,
a further useful choice for the mapping between $J$ and $L$
can be obtained by associating the parameter $v_{L}$
with the coefficient $r_{J}$ of the $N$-qubit ket
in the order of growing number of excitations.
Let us recall that, for an arbitrary $N$-qubit state,
the global ground state $|00\ldots 0\rangle$ is labeled by the index $J=0$;
the kets with single excitation, e.g. of the form $|010\ldots 0\rangle$,
are labeled by the indices $J=2^{p}$ with $p=0,\ldots,N-1$;
the kets with two excitations, e.g. of the form $|01010\ldots 0\rangle$,
are labeled by the indices
$J=2^{q}+2^{s}$ with $q=1,\ldots,N-1$ and $s=0,\ldots, p-1$;
and so on. One can decide to choose the following mapping for the parametrization:
\begin{eqnarray}
&& r_{0} = v_{0} \,, \nonumber \\
&& r_{2^{p}} = v_{p+1} \,, \qquad p=0,\ldots,N-1 \,, \nonumber \\
&& r_{2^{q}+2^{s}} = v_{N+q+s} \,, \nonumber \\
&&\quad q=1,\ldots,N-1 \,, s=0,\ldots, q-1 \,, \nonumber \\
&& \ldots\ldots\ldots \,. \label{orderexcit}
\end{eqnarray}
Clearly, the choice of the mapping between $r_{J}$ and $v_{L}$
can be suitably done by looking at the terms that survive 
in the squared overlap (\ref{overlap2rel}).

\section{Evaluation of Eqs.~(\ref{2sepruleW}), (\ref{KsepruleW}),
(\ref{overlapWGHZN}), and (\ref{KsepruleWlike}) }
\label{AppendixB}

Here we outline the analytical evaluation of the squared
overlaps (\ref{2sepruleW}), (\ref{KsepruleW}),
(\ref{overlapWGHZN}), and (\ref{KsepruleWlike}).
The quantity $\Lambda_2^2(M_1|M_2)$ in Eq.~(\ref{2sepruleW})
involves only coefficients of the form
$r_{0}^{(s)}$ and $r_{2^{p}}^{(s)}$ $(s=1,2)$.
Therefore, when using the hyperspherical representation
(\ref{hypersphparam}), by exploiting the freedom in the ordering,
we choose the mapping given in Eq.~(\ref{orderexcit}), specifically
\begin{eqnarray}
&&r_{0}^{(s)} \,=\, v_{0}^{(s)} \,=\, \cos\delta_{0}\,, \nonumber \\
&&r_{2^{p}}^{(s)} \,=\, v_{p+1}^{(s)} \,=\,
\sin\delta_{0}^{(s)}\cdots \sin\delta_{p}^{(s)}\cos \delta_{p+1}^{(s)} \,, \nonumber \\
&& p=0,\ldots,M_{s}-1 \,.
 \label{mappingW}
\end{eqnarray}
Such parametrization leads to the explicit expression
\begin{eqnarray}
&&\Lambda_{2}^{2}(M_1|M_2) = \max_{\{ \delta_{J_{s}}^{(s)} \}}
\frac{1}{N}
\{\cos\delta_{0}^{(1)} \sin \delta_{0}^{(2)} [\cos\delta_{1}^{(2)} \nonumber \\
&& \nonumber \\
&& +
\sin\delta_{1}^{(2)}[\cos\delta_{2}^{(2)} + \sin\delta_{2}^{(2)} [\ldots
[\cos \delta_{M_2 -1}^{(2)} + \sin \delta_{M_2 -1}^{(2)} \times \nonumber \\
&& \nonumber \\
&& \times \cos\delta_{M_2}^{(2)}]]]]+
\cos \delta_{0}^{(2)} \sin\delta_{0}^{(1)}[\cos\delta_{1}^{(1)}
+ \sin \delta_{1}^{(1)} \times
\nonumber \\
&& \nonumber \\
&& \times [\cos\delta_{2}^{(1)} +  \sin\delta_{2}^{(1)} [\ldots  [\cos
\delta_{M_1 -1}^{(1)} + \sin \delta_{M_1 -1}^{(1)} \times \nonumber \\
&& \nonumber \\
&& \times \cos \delta_{M_1}^{(1)}]]]]\}^{2}
\; .  \label{2sepEMWN}
\end{eqnarray}
Analyzing the structure of the $|W^{(N)}\rangle$ states, it is
convenient, given a generic integer $M$, and a set of generic
variables $\delta_{i}, \quad i=1,\ldots,M$, to introduce the following
function:
\begin{eqnarray}
&&f(\delta_{1}^{(s)},\ldots,\delta_{M_{s}}^{(s)}) = \cos\delta_{1}^{(s)}+\sin\delta_{1}^{(s)}
 [\cos\delta_{2}^{(s)} + \sin\delta_{2}^{(s)}\times \nonumber \\
&& \nonumber \\
&& \times [\ldots [\cos\delta_{M_{s}-1}^{(s)}+ \sin\delta_{M_{s}-1}^{(s)}
\cos\delta_{M_{s}}^{(s)}]]] .
\label{maxfuncW}
\end{eqnarray}
By means of Eq.~(\ref{maxfuncW}), the expression (\ref{2sepEMWN})
can be recast in the more compact form
\begin{eqnarray}
&&\Lambda_{2}^{2}(M_1|M_2) = \max_{\{ \delta_{J_{s}}^{(s)} \}}
\frac{1}{N}
\{\cos\delta_{0}^{(1)} \sin\delta_{0}^{(2)} \times
\nonumber \\
&& \nonumber \\
&&\times f(\delta_{1}^{(2)}, \ldots, \delta_{M_2}^{(2)})
+ \cos\delta_{0}^{(2)} \sin \delta_{0}^{(1)} \times \nonumber \\
&& \nonumber \\
&& \times f(\delta_{1}^{(1)}, \ldots, \delta_{M_1}^{(1)}) \}^{2} \; .
\label{2sepEMWNbis}
\end{eqnarray}
To proceed, we first maximize the function $f(\delta_{1}^{(s)},\ldots,\delta_{M}^{(s)})$,
i.e. Eq.~(\ref{maxfuncW}), over the $M_{s}$ independent
variables $\delta_{h}^{(s)}$ $(h=1,\ldots,M_{s})$.
This task can be accomplished as follows:
First, trivially, $\delta_{M}^{(s)}=0$ maximizes $\cos \delta_{M}^{(s)}$.
Next, the contribution $(\cos\delta_{M-1}^{(s)}+ \sin\delta_{M-1}^{(s)})$ reaches
the maximum value $\sqrt{2}$ for $\delta_{M-1}^{(s)}=\frac{\pi}{4}$.
After the elimination of the parameters $\delta_{M-1}^{(s)}$ and $\delta_{M}^{(s)}$,
the term $(\cos\delta_{M-2}^{(s)}+ \sin\delta_{M-2}^{(s)}\sqrt{2})$ appears.
Observing that terms of the form $(\cos \theta+ \sin \theta \sqrt{n})$
acquire the maximum value $\sqrt{1+ n}$ for $\theta=\arcsin\sqrt{\frac{n}{1+n}}$,
the cascade maximization procedure yields that Eq.~(\ref{maxfuncW}) is maximized
at the value $\sqrt{M}_{s}$ for $\delta_{M-h}^{(s)} = \arcsin\sqrt{\frac{h}{1+h}}$,
with $h=0,1,\ldots,M-1$.
Reminding that $1 \leq M_1 \leq M_2 = N - M_1$, and performing the
final maximization in Eq.~(\ref{2sepEMWN}) yields
\begin{equation}
\label{MaxOverlap}
\Lambda_{2}^{2}(M_1|M_2) = \frac{M_2}{N} \; .
\end{equation}
The above maximum overlap squared is reached for the values
$\delta_{0}^{(1)} = 0$, $\delta_{0}^{(2)} = \frac{\pi}{2}$,
$\delta_{N-M-h}^{(2)} = \arcsin \sqrt{\frac{h}{h+1}}$ with $h = 0, 1,
\ldots, M_2 -1$. \\
One can now consider the squared overlap $\Lambda_{K}^{2}(M_1 |\ldots |M_K)$,
Eq.~(\ref{KsepruleW}).
Proceeding as for Eq.~(\ref{2sepruleW}), and thus using,  for any $s$,
the hyperspherical representation (\ref{hypersphparam})
with the mapping given by Eq.~(\ref{mappingW}),
one can express Eq.~(\ref{KsepruleW}) in terms of the angular parameters
$\delta_{J_{s}}^{(s)}$.
Moreover, by using definition (\ref{maxfuncW}),
\begin{eqnarray}
&& \Lambda_{K}^{2}(M_1 |\ldots |M_K) = \nonumber \\
&& \nonumber \\
&&\max_{ \{ \delta_{J_{s}}^{(s)} \} }
\frac{1}{N} \Big[ \cos\delta_{0}^{(1)} \cos\delta_{0}^{(2)} \cdots \cos \delta_{0}^{(K-1)} \times
\nonumber \\
&& \nonumber \\
&& \times \sin\delta_{0}^{(K)} f(\delta_{1}^{(K)},\ldots,\delta_{M_{K}}^{(K)}) +
\cos\delta_{0}^{(1)} \cos\delta_{0}^{(2)} \cdots \times
\nonumber \\
&& \nonumber \\
&& \times \cos\delta_{0}^{(K-2)} \sin\delta_{0}^{(K-1)} f(\delta_{1}^{(K-1)}, \ldots,
\delta_{M_{K-1}}^{(K-1)}) \cos\delta_{0}^{(K)} \nonumber \\
&& \nonumber \\
&& + \ldots + \sin\delta_{0}^{(1)} f(\delta_{1}^{(1)},\ldots,\delta_{M_{1}}^{(1)}) \times
\nonumber \\
&& \nonumber \\
&& \times \cos \delta_{0}^{(2)} \cdots \cos \delta_{0}^{(K-1)} \cos \delta_{0}^{(K)} \Big]^{2} \; .
\label{KsepruleWbis}
\end{eqnarray}
The maximization of the functions $f$, corresponding to the replacement
$f(\delta_{1}^{(s)},\ldots,\delta_{M_{s}}^{(s)}) \rightarrow \sqrt{M_{s}}$,
reduces relation (\ref{KsepruleWbis}) to Eq.~(\ref{KsepruleWtris}). \\
Next, we consider the squared overlap (\ref{overlapWGHZN}).
By choosing the following mapping
\begin{eqnarray}
&&r_{0}^{(s)} \,=\, v_{0}^{(s)} \,=\, \cos\delta_{0}^{(s)} \,, \nonumber \\
&& r_{2^{M_{s}}-1}^{(s)} \,=\, v_{1}^{(s)} \,=\, \sin\delta_{0}^{(s)}\cos\delta_{1}^{(s)}\,, \nonumber \\
&&r_{2^{p}}^{(s)} \,=\, v_{p+2}^{(s)} \,=\,
\sin\delta_{0}^{(s)}\cdots \sin\delta_{p+1}^{(s)}\cos \delta_{p+2}^{(s)} \,, \nonumber \\
&& p=0,\ldots,M_{s}-1 \,,
 \label{mappingWGHZ}
\end{eqnarray}
we obtain the relation:
\begin{eqnarray}
&&\Lambda_{2}^{2}(M_{1}|M_{2}) = \max_{\{ \delta_{q}^{(s)} \}}
\Big|\frac{\cos\eta}{\sqrt{N}} \Big(\cos\delta_{0}^{(1)}\sin\delta_{0}^{(2)}\sin\delta_{1}^{(2)} \times
\nonumber \\
&& \nonumber \\
&& f(\delta_{2}^{(2)},\ldots,\delta_{M_{2}+1}^{(2)}) + \sin\delta_{0}^{(1)}\sin\delta_{1}^{(1)}\sin\delta_{0}^{(2)}
\times
\nonumber  \\
&& \nonumber \\
&&f(\delta_{2}^{(1)},\ldots,\delta_{M_{1}+1}^{(1)}) \Big)+\frac{\sin\eta}{\sqrt{2}} \Big( \cos\delta_{0}^{(1)} \cos\delta_{0}^{(2)}+ \sin\delta_{0}^{(1)} \times \nonumber \\
&& \nonumber \\
&&\cos\delta_{1}^{(1)}
\sin\delta_{0}^{(2)}\cos\delta_{1}^{(2)} \Big)
\Big|^{2} \,.
\label{overlapWGHZN3}
\end{eqnarray}
By maximizing the functions $f$, we arrive at the final expression Eq.~(\ref{overlapWGHZN2}). \\
Coming to the squared overlap $\Lambda_{K}^{2}(Q_1 |\ldots |Q_K)$,
i.e. Eq.~(\ref{KsepruleWlike}), one can
proceed as for Eq.~(\ref{KsepruleW}) in order to obtain a relation identical to
Eq.~(\ref{KsepruleWbis}) with the functions $f$ replaced by the functions $F$ defined as:
\begin{eqnarray}
&&F(\delta_{1}^{(s)},\ldots,\delta_{M_{s}}^{(s)}) =
\Gamma_{1}^{(s)} \cos\delta_{1}^{(s)}+\sin\delta_{1}^{(s)}
[\Gamma_{2}^{(s)} \cos\delta_{2}^{(s)} \nonumber \\
&& \nonumber \\
&& + \sin\delta_{2}^{(s)} [\ldots [\Gamma_{M_{s}-1}^{(s)}\cos\delta_{M_{s}-1}^{(s)}+ \sin\delta_{M_{s}-1}^{(s)} \times \nonumber \\
&& \nonumber \\
&&\times \Gamma_{M_{s}}^{(s)} \cos\delta_{M_{s}}^{(s)}]]] \;,
\label{maxfuncWlike}
\end{eqnarray}
where $\Gamma_{p+1}^{(s)} = \gamma_{p+1+M_{s+1}+\ldots + M_{K}}$,
with $p=0,\ldots , M_{s}-1$. It is straightforward to notice
that the following relation holds:
\begin{equation}
\max_{\delta} \{x\cos\delta + y\sin\delta\} \,=\, \sqrt{x^{2}+y^{2}}  \,,
\label{maximizrelation}
\end{equation}
where $x$ and $y$ are real parameters.
By exploiting iteratively the maximization procedure to the terms of this form,
we find that Eq.~(\ref{maxfuncWlike}) is maximized at the value
$\sqrt{\sum_{p=0}^{M_{s}-1}\Big(\Gamma_{p+1}^{(s)}\Big)^{2}}$.
Finally, maximizing all the functions $F$, the squared overlap (\ref{KsepruleWlike})
reduces to Eq.~(\ref{KsepruleWlike2}).

\section{Evaluation of Eq.~(\ref{overlap2sepMagnon})}
\label{AppendixMagnon}

In order to compute the squared overlap $\Lambda_{2}^{2}(M_{1}|M_{2})$,
Eq.~(\ref{overlap2sepMagnon}), we exploit again the hyperspherical coordinates
(\ref{hypersphparam}), and we introduce the following mapping
\begin{eqnarray}
&&r_{0}^{(s)} \,=\, v_{0}^{(s)} \,=\, \cos\delta_{0}^{(s)} \,, \nonumber \\
&& r_{2^{p}+2^{q}}^{(s)} \,=\, v_{l(p,q)+1}^{(s)} \,=\, \sin\delta_{0}^{(s)}\cdots \sin\delta_{l(p,q)}^{(s)}\cos \delta_{l(p,q)+1}^{(s)}\,, \nonumber \\
&& l(p,q) = 0, \ldots, L_{s}-1 \,, \;\; p=1,\ldots,M_{s}-1 \,, \nonumber \\
&& q=0,\ldots,p-1 \,, \nonumber \\
&&r_{2^{p}}^{(s)} \,=\, v_{p+1+L_{s}}^{(s)} \,=\,
\sin\delta_{0}^{(s)}\cdots \sin\delta_{p+L_{s}}^{(s)}\cos \delta_{p+1+L_{s}}^{(s)} \,, \nonumber \\
&& p=0,\ldots,M_{s}-1 \,,
\label{mappingMagnon}
\end{eqnarray}
where $L_{s}= \left( \begin{array}{c}
                M_{s} \\
                2
              \end{array} \right) = \frac{M_{s}(M_{s}-1)}{2}$.
Furthermore, by defining the function:
\begin{equation}
g(\delta_{1}^{(s)},\ldots, \delta_{L_{s}}^{(s)}) \,=\,
\sin\delta_{0}^{(s)}\cdots \sin\delta_{L_{s}}^{(s)} \,,
\label{gmagnon}
\end{equation}
Eq.~(\ref{overlap2sepMagnon}) is recast in the form
\begin{eqnarray}
&&\Lambda_{2}^{2}(M_{1}|M_{2}) \,=\,
\max_{\delta_{J_{s}}^{(s)}}
\Big|\sin\delta_{0}^{(1)} \cos\delta_{0}^{(2)} f(\delta_{1}^{(1)},\ldots,\delta_{L_{1}}^{(1)})
\nonumber \\
&& \nonumber \\
&& + \cos\delta_{0}^{(1)}\sin\delta_{0}^{(2)} f(\delta_{1}^{(2)},\ldots,\delta_{L_{2}}^{(2)})
+ \sin\delta_{0}^{(1)}\sin\delta_{0}^{(2)} \times \nonumber \\
&& \nonumber \\
&& \times g(\delta_{1}^{(1)},\ldots,\delta_{L_{1}}^{(1)}) \,
f(\delta_{1+L_{1}}^{(1)},\ldots,\delta_{M_{1}+L_{1}}^{(1)}) \times \nonumber \\
&& \nonumber \\
&& \times g(\delta_{1}^{(2)},\ldots,\delta_{L_{2}}^{(2)})
f(\delta_{1+L_{2}}^{(2)},\ldots,\delta_{M_{2}+L_{2}}^{(2)}) \Big|^{2} \,.
\label{overlap2sepMagnon2}
\end{eqnarray}
The functions $f(\delta_{1+L_{s}}^{(s)},\ldots,\delta_{M_{s}+L_{s}}^{(s)})$
in the last term of Eq.~(\ref{overlap2sepMagnon2}) can be maximized at the
value $\sqrt{M_{s}}$.
It is quite straightforward to observe that the functions $f$ and $g$ are
connected by the relation
\begin{eqnarray}
f(\delta_{1}^{(s)},\ldots,\delta_{L_{s}}^{(s)})= &&
f(\delta_{1}^{(s)},\ldots,\delta_{L_{s}-1}^{(s)}) \nonumber \\
+&& g(\delta_{1}^{(s)},\ldots,\delta_{L_{s}-1}^{(s)}) \cos\delta_{L_{s}}.
\label{fgconnect}
\end{eqnarray}
By using property (\ref{fgconnect}) and applying the maximization
relation (\ref{maximizrelation}) to the variables $\delta_{L_{1}}^{(1)},\delta_{L_{1}-1}^{(1)},\ldots,\delta_{1}^{(1)},\delta_{0}^{(1)}$,
Eq.~(\ref{overlap2sepMagnon2}) reduces to
\begin{eqnarray}
&&\Lambda_{2}^{2}(M_{1}|M_{2}) \,=\,
\max_{\delta_{J_{2}}^{(2)}} \left( \begin{array}{c}
                                     N \\
                                     2
                                   \end{array} \right)^{-1}
\Big\{ \cos^{2}\delta_{0}^{(2)} L_{1} + \sin^{2}\delta_{0}^{(2)} \times \nonumber \\
&& \nonumber \\
&&\times \big[  f^{2}(\delta_{1}^{(2)},\ldots,\delta_{L_{2}}^{(2)}) +
g^{2}(\delta_{1}^{(2)},\ldots,\delta_{L_{2}}^{(2)}) M_{1} M_{2}\big] \Big\} .
\label{overlap2sepMagnon3}
\end{eqnarray}
Observing that $L_{1} < L_{2},M_{1} M_{2}$, a further maximization on
the variable $\delta_{0}^{(2)}$ yields
\begin{eqnarray}
&&\Lambda_{2}^{2}(M_{1}|M_{2}) \,=\,
\max_{\delta_{J_{2}}^{(2)}} \left( \begin{array}{c}
                                     N \\
                                     2
                                   \end{array} \right)^{-1}
\big[  f^{2}(\delta_{1}^{(2)},\ldots,\delta_{L_{2}}^{(2)}) \nonumber \\
&& \nonumber \\
&&+
g^{2}(\delta_{1}^{(2)},\ldots,\delta_{L_{2}}^{(2)}) M_{1} M_{2}\big]  \,.
\label{overlap2sepMagnon4}
\end{eqnarray}
In order to maximize Eq.~(\ref{overlap2sepMagnon4}), one has to find the conditions
for the vanishing of the first partial derivatives with respect to each variable $\delta_{q}^{(2)}$
with $q=1,\ldots,L_{2}$, and to select the absolute maximum in the intervals
$0\leq \delta_{q}^{(2)}\leq \frac{\pi}{2}$.
Without reporting the whole set of the calculations,
we only outline briefly the main steps of the maximization procedure.
The first partial derivative with respect to $\delta_{L_{2}}^{(2)}$ yields the following condition:
\begin{eqnarray}
&&g(\delta_{1}^{(2)},\ldots,\delta_{L_{2}}^{(2)})
[f(\delta_{1}^{(2)},\ldots,\delta_{L_{2}}^{(2)}) \nonumber \\
&& \nonumber \\
&&- M_{1} M_{2} \, g(\delta_{1}^{(2)},\ldots,\delta_{L_{2}-1}^{(2)}) \, \cos\delta_{L_{2}}^{(2)}] =0 \,.
\label{firstderivMagn}
\end{eqnarray}
By using Eq.~(\ref{firstderivMagn}) together with all the other partial derivatives
with respect to $\delta_{q}^{(2)}$ $(q=1,\ldots,L_{2}-1)$,
it is not difficult to verify that the absolute maximum of $\Lambda_{2}^{2}(M_{1}|M_{2})$
is given by
\begin{equation}
\Lambda_{2}^{2}(M_{1}|M_{2}) \,=\,
\left( \begin{array}{c}
                                     N \\
                                     2
                                   \end{array} \right)^{-1}
\max \{ L_{2} \,,\,  M_{1} M_{2} \}  \,.
\label{overlap2sepMagnon5}
\end{equation}
Indeed, from Eq.~(\ref{firstderivMagn}) one has that if
$g(\delta_{1}^{(2)},\ldots,\delta_{L_{2}}^{(2)}) = 0 $, then
Eq.~(\ref{overlap2sepMagnon4}) reduces to the maximization of the function
$f(\delta_{1}^{(2)},\ldots,\delta_{L_{2}}^{(2)})$, i.e. $\sqrt{L_{2}}$.
On the other hand, if $$f(\delta_{1}^{(2)},\ldots,\delta_{L_{2}}^{(2)}) = M_{1} M_{2} \, g(\delta_{1}^{(2)},\ldots,\delta_{L_{2}-1}^{(2)}) \, \cos\delta_{L_{2}},$$
then the maximum is reached for $\delta_{q}=\frac{\pi}{2}$,
i.e. $f(\frac{\pi}{2},\ldots,\frac{\pi}{2})= 0$ and $g(\frac{\pi}{2},\ldots,\frac{\pi}{2})= 1$.
It can be verified that all the other values of $\delta_{q}$ leading to the
vanishing of the first partial derivatives are associated with relative maxima.

\acknowledgments

We acknowledge MIUR under PRIN National Project 2005,
CNR-INFM Research and Development Center {\it "Coherentia"}, INFN,
and ISI Foundation for financial support.


\begin{thebibliography}{99}

\bibitem{Einstein} A. Einstein, B. Podolsky, and N. Rosen, Phys. Rev.
{\bf 47}, 777, (1935).

\bibitem{Schroedinger} E. Schroedinger, Naturwiss. {\bf 23}, 807
(1935); {\it ibid.} {\bf 23}, 823 (1935); {\it ibid.} {\bf 23},
844 (1935).

\bibitem{Bennett1} C. H. Bennett, H. J. Bernstein, S. Popescu, and B. Schumacher,
Phys. Rev. A {\bf 53}, 2046 (1996).

\bibitem{Bennett2} C. H. Bennett, D. P. DiVincenzo, J. A. Smolin, and W. K. Wootters,
Phys. Rev. A {\bf 54}, 3824 (1996).

\bibitem{Vidal} G. Vidal, J. Mod. Opt. {\bf 47}, 355 (2000).

\bibitem{PlenioVirmani} For recent comprehensive reviews on various aspects of
entanglement theory, see: M. Plenio and S. Virmani, Quant. Inf. Comp. {\bf 7}, 1 (2007);
I. Bengtsson and K. Zyczkowski, {\it Geometry of Quantum States} (Cambridge University
Press, 2006); R. Horodecki, P. Horodecki, M. Horodecki, and
K. Horodecki, quant-ph/0702225.

\bibitem{Popescu} S. Popescu and D. Rohrlich, Phys. Rev. A {\bf 56}, R3319 (1997).

\bibitem{EntRelEntr}
V. Vedral and M. B. Plenio, Phys. Rev. A {\bf 57}, 1619 (1998).

\bibitem{Negativity}
G. Vidal and R. F. Werner, Phys. Rev. A {\bf 65}, 032314 (2002).

\bibitem{HillWootters} S. Hill and W. K. Wootters,  Phys. Rev. Lett. {\bf 78}, 5022 (1997).

\bibitem{Wootters} W. K. Wootters, Phys. Rev. Lett. {\bf 80}, 2245 (1998).

\bibitem{SLOCC} C. H. Bennett, S. Popescu, D. Rohrlich, J. A. Smolin, and
A. V. Thapliyal, Phys. Rev. A {\bf 63}, 012307 (2001).

\bibitem{2diffwayent} W. D\"{u}r, G. Vidal, and J. I. Cirac,
Phys. Rev. A {\bf 62}, 062314 (2000).

\bibitem{9diffwayent} F. Verstraete, J. Dehaene, B. De Moor, and H. Verschelde,
 Phys. Rev. A {\bf 65}, 052112 (2002).

\bibitem{GHZst} D. M. Greenberger, M. Horne, and A. Zeilinger,
in {\it Bell's Theorem, Quantum Theory, and Conceptions of the
Universe}, M. Kafatos Ed. (Kluwer, Dordrecht, 1989).

\bibitem{CoffKundWoot} V. Coffman, J. Kundu, and W. K. Wootters,
Phys. Rev. A {\bf 61}, 052306 (2000).

\bibitem{EisertBriegel} J. Eisert and H. J. Briegel, Phys. Rev. A {\bf 64}, 022306 (2001).

\bibitem{Wallach} D. A. Meyer and N. R. Wallach, J. Math. Phys. {\bf 43}, 4273 (2002).

\bibitem{Brennen} G. K. Brennen, Quantum Inf. Comput. {\bf 3}, 619 (2003).

\bibitem{Scott} A. J. Scott, Phys. Rev. A {\bf 69}, 052330 (2004).

\bibitem{Oliveira} T. R. de Oliveira, G. Rigolin, and M. C. de Oliveira,
Phys. Rev. A {\bf 73}, 010305(R) (2006); G. Rigolin, T. R. de
Oliveira, and M. C. de Oliveira, Phys. Rev. A {\bf 74}, 022314
(2006).

\bibitem{Pascazio} P. Facchi, G. Florio, and S. Pascazio,
Phys. Rev. A {\bf 74}, 042331 (2006).

\bibitem{multirelentropy} M. B. Plenio and V. Vedral, J. Phys. A
{\bf 34}, 6997 (2001).

\bibitem{GMEShimony} A. Shimony, Ann. N. Y. Acad. Sci. {\bf 755}, 675
(1995).

\bibitem{GMEBarnumLind} H. Barnum and N. Linden, J. Phys. A: Math. Gen. {\bf 34}, 6787
(2001).

\bibitem{GMEWeiGold} T.-C. Wei and P. M. Goldbart, Phys. Rev. A {\bf 68}, 042307
(2003).

\bibitem{EisertGross} J. Eisert and D. Gross, in {\it Lectures on quantum information},
D. Bruss and G. Leuchs Eds. (Wiley-VCH, Weinheim, 2006), and
quant-ph/0505149.

\bibitem{Cavalcanti} D. Cavalcanti, Phys. Rev. A {\bf 73}, 044302
(2006).

\bibitem{Guehne} O. G\"uhne, M. Reimpell, and R. F. Werner,
Phys. Rev. Lett. {\bf 98}, 110502 (2007).

\bibitem{Eisert} J. Eisert, F. G. S. L. Brand\~{a}o, and K. M. R. Audenaert,
New J. Phys. {\bf 9}, 46 (2007).

\bibitem{GMEBoundEntang} T.-C. Wei, J. B. Altepeter, P. M. Goldbart, and W. J. Munro,
Phys. Rev. A {\bf 70}, 022322 (2004).

\bibitem{GMEQpts} T.-C. Wei, D. Das, S. Mukhopadyay, S. Vishveshwara, and P. M. Goldbart,
Phys. Rev. A {\bf 71}, 060305(R) (2005).

\bibitem{GMEOrus} R. Or\'{u}s, Phys. Rev. Lett. {\bf 100}, 130502 (2008).

\bibitem{GMELmg} R. Or\'{u}s, S. Dusuel, and J. Vidal, arXiv:0803.3151v1.

%%\bibitem{GMEGeneralizations} Y. Cao and A. M. Wang, J. Phys. A: Math. Theor. {\bf 40}, 3507 (2007);
%%D. Ostapchuk, G. Passante, R. Kobes, and G. Kunstatter, arXiv:0707.4020 [quant-ph].

%%\bibitem{GMEMethods} L. Tamaryan, D. K. Park and S. Tamaryan, Phys. Rev. A {\bf 77}, 022325 (2008);
%%L. Tamaryan, D. Park, J.-W. Son, S. Tamaryan, arXiv:0803.1040 [quant-ph].

\bibitem{BrieRaus} H. J. Briegel and R. Raussendorf, Phys. Rev. Lett. {\bf 86}, 910 (2001).

\bibitem{4qubitClusterGen} N. Kiesel, C. Schmid, U. Weber, G.
T\'{o}th, O. G\"{u}hne, R. Ursin, and H. Weinfurter, Phys. Rev.
Lett. {\bf 95}, 210502 (2005).

\bibitem{Report} F. Dell'Anno, S. De Siena, and F. Illuminati,
Phys. Rep. {\bf 428}, 53 (2006).

\bibitem{1wayQC} R. Raussendorf and H. J. Briegel,
Phys. Rev. Lett. {\bf 86}, 5188 (2001).

\bibitem{Mattis} D. C. Mattis, {\em The Theory of Magnetism I - Statics and Dynamics}
(Springer-Verlag, Berlin, 1988).

\bibitem{ManyMagnons} T. Morimae, A. Sugita, and A. Shimizu, Phys. Rev. A {\bf 71}, 032317 (2005).

\bibitem{GeneralWstates}
P. Agrawal, and A. K. Pati, Phys. Rev. A {\bf 74}, 062320 (2006);
S. Adhikari and S. Gangopadhyay, arXiv:0803.0607 [quant-ph].

\bibitem{WstatePrepar} G. C. Guo, and Y. -S. Zhang, Phys. Rev. A {\bf 65}, 054302 (2002);
V. N. Gorbachev, A. A. Rodichkina, A. I. Trubilko, and A. I. Zhiliba,
Phys. Lett. A {\bf 310}, 339 (2003);
A. Biswas and G. S. Agarwal, J. Mod. Opt. {\bf 51}, 1627 (2004).

%%\bibitem{Brody} D. C. Brody, A. C. T. Gustavsson, and L. P.
%%Hughston, J. Phys.: Conf. Ser. {\bf 67}, 012044 (2007).

\end{thebibliography}
\end{document}